\documentstyle[dina4,epsfig,rotate,11pt]{article}
\newcommand{\be}{\begin{equation}}   
\newcommand{\ee}{\end{equation}}   
\newcommand{\bea}{\begin{eqnarray}}   
\newcommand{\eea}{\end{eqnarray}}   
\newcommand{\fn}{$f_N(\vec{r},\,\vec{p},\,t)\, \, $}

\newcommand{\non}{\nonumber}

\begin{document}

\title{Pion-Production in Heavy-Ion Collisions at SIS 
energies\footnote{Supported by BMBF and GSI Darmstadt} {}\footnote{part of the
PhD thesis of S. Teis} }
\author{S. Teis, W. Cassing, M. Effenberger, A. Hombach, U. Mosel and 
	  Gy. Wolf\\
	Institut f\"{u}r Theoretische Physik, Universit\"{a}t Giessen \\
	D-35392 Giessen, Germany}
\maketitle
\begin{abstract}
We investigate the production of pions in heavy-ion collisions in the energy
range of $1$ - $2$ GeV/A. The dynamics of the nucleus-nucleus 
collisions is described by a set of coupled
transport equations of the Boltzmann-Uehling-Uhlenbeck type for baryons and
mesons. Besides the $N(938)$ and the $\Delta(1232)$ we also take into account
nucleon resonances up to masses of $1.9 \,\,
GeV/c^2$ as well as $\pi$-, $\eta$- and  $\rho$-mesons. We study in detail
the influence of the higher baryonic resonances and the 
$2\pi$-production channels ($NN\to NN \pi\pi$) on the pion spectra 
in comparison to $\pi^-$ data from $Ar + KCl$ collisions at
$1.8$ GeV/A and $\pi^0$-data for $Au+Au$ 
at 1.0 GeV/A. We, furthermore, present a detailed comparison of differential
pion angular distributions with the BEVALAC data for Ar + KCl at 1.8 GeV/A. 
The general agreement obtained indicates that the overall reactions 
dynamics is well described by our novel transport approach.
\end{abstract}

\vspace{2cm}
\begin{section}{Introduction}
Relativistic heavy-ion collisions provide a unique possibility to study
nuclear matter at high density and far away from equilibrium. 
During the course of a
heavy-ion collision at $1$ - $2$ GeV/A the nuclear matter
is compressed up to $2 - 3$ times normal nuclear matter density $\rho_0$
before it expands again. The experimental probes that provide information
about the compressed stage of the reaction are collective observables such
as flow patterns or differential spectra of particles that have been
additionally  produced during the collision.

Over the last 10 years transport theories as BUU 
\cite{bertsch88,cassing88,cassing90}
and QMD, IQMD \cite{aichelin91,bass93,bass1,bass2} 
have been very successful in describing the
reaction dynamics of heavy-ion collisions. It has been found that the
experimental meson spectra can be well understood when assuming the
excitation and 
subsequent decay of nucleon resonances during the compressed stage.
Since pions couple strongly to these resonances, the differential pion spectra
provide a well suited probe for the dynamics of baryonic resonances. 
Furthermore, due to the
low production threshold, pions are produced and reabsorbed quite frequently
and thus provide a signal for the whole dynamical evolution of the heavy-ion 
reaction.
This implies in particular that a transport theoretical description of
heavy-ion collisions has to reproduce the pion yields correctly before one
can draw any further conclusions on more specific channels from the model. 
Especially for observables like dileptons the pion annihilation 
($\pi^+ \pi^- \to e^+e^-$) plays an important role; also the production of kaons 
(via $\pi N \to K \Lambda$) and $\eta$-mesons (via $\pi N \to \eta N$) is
strongly influenced by the pion induced channels. 

In this work we present a new implementation of the BUU-model,
denoted by Coupled-Channel-BUU (CBUU), which 
as an extension to previous realisations in \cite{wolf90,wolf93} - or
compared to the transport model IQMD \cite{bass93,bass1,bass2} - includes 
all baryonic resonances up to masses of
$1950$ $MeV/c^2$. The higher resonances are expected to contribute to
the low as well as to the high momentum regimes of pion spectra via 1$\pi$
and 2$\pi$ decay channels, respectively. Apart from
the pions ($\pi^+, \, \pi^0,\, \pi^-$) we now also explicitly propagate 
$\eta$'s and $\rho$-mesons ($\rho^+, \, \rho^0,\, \rho^-$) as 
mesonic degrees of freedom thus including hadronic excitations up to
about 1 GeV of excitation energy. 
After introducing our model in section 2 and the various elastic 
and inelastic cross
sections in section 3 we present detailed comparisons with $\pi^-$data
taken at the BEVALAC \cite{stock86} and the SIS in section 4. A summary and
discussion of open problems concludes the paper in section 5.
\end{section}
\begin{section}{The CBUU-Model}
\begin{subsection}{Basic Equations}
In line with refs. \cite{bertsch88,cassing88,cassing90,kweber93} 
the dynamical evolution of heavy-ion collisions or 
hadron-nucleus reactions below the pion-production threshold is described  
by a transport equation for the nucleon one-body 
phase-space distribution function $f_N(\vec{r},\vec{p},t)$, 
\bea
\label{buueq}
\frac{\partial f(\vec{r},\,\vec{p},\,t)}{\partial t} &+& \left\{ \frac{\vec{p}}{E} + 
\frac{m^*(\vec{r},\vec{p})}{E}\, \vec{\nabla}_p\, U(\vec{r},\, \vec{p}) \right\} 
\, \vec{\nabla}_r f(\vec{r},\,\vec{p},\,t)  \nonumber  \\
& &+ \left\{ -\frac{m^*(\vec{r},\vec{p})}{E}\, \vec{\nabla}_r  
U(\vec{r},\, \vec{p}) \right\}\, \vec{\nabla}_r f(\vec{r},\,\vec{p},\,t) 
\,\, =\, \, I_{coll}[f(\vec{r},\,\vec{p},\,t)],
\eea
where $\vec{r}$ and $\vec{p}$ denote the spatial and the momentum coordinate of 
the nucleon, 
respectively, while $N$ stands for a proton ($p$) or neutron ($n$). 
The effective mass $m^*(\vec{r}, \, \vec{p})$ in 
eq. (\ref{buueq}) includes the nucleon restmass $m_N$ ($ = 938$ MeV/c) 
as well as a scalar momentum-dependent mean field potential 
$U(\vec{r},\, \vec{p})$, 
\be
\label{effmass}
m^*(\vec{r}, \, \vec{p})  = m_N + U(\vec{r},\, \vec{p}).
\ee
The nucleon quasi-particle properties then read as 
\be
\label{esingle}
E = \sqrt{ m^*( \vec{r}, \, \vec{p})^2 + \vec{p}^2}.
\ee
Physically the l.h.s. of eq. (\ref{buueq}) represents the Vlasov equation for a gas of
non-interacting nucleons moving in the scalar momentum-dependent mean-field potential
$U(\vec{r},\, \vec{p})$. The Vlasov equation as given above
can be derived from a manifest covariant transport equation using scalar and
vector self-energies depending on the four-momentum ($p^\mu = (E,\, \vec{p}))$
of the particles \cite{kweber93,tomo} when neglecting the vector self-energies.
Due to the quasiparticle mass-shell constraint given by (\ref{esingle})
the scalar potential effectively depends only on the three-momentum of the nucleons as 
($U(\vec{p},E(\vec{p})$).\\
The r.h.s. of the BUU-equation (i.e. the collision integral
$I_{coll}(f(\vec{r},\,\vec{p},\,t))$ \cite{bertsch88,cassing88}) describes 
the time evolution of \fn       due to two-body
collisions among the nucleons. For example, the alteration in the one-body 
phase-space distribution function $f(\vec{r}_1, \, \vec{p}_1,\, t)$  
due to the elastic scattering of two 
nucleons 
\bea 
N_1 + N_2 \longleftrightarrow N_3 + N_4, \nonumber 
\eea
with momenta ($\vec{p}_i, \, \, i = 1, .., 4$) is given 
by \cite{cassing88,cassing90} 
\bea
\label{collint}
I_{coll}\left[ f_1(\vec{r},\,\vec{p}_1,\,t)\right] & = & \sum_{2,3,4}
\frac{g}{(2\pi)^3} \, 
\int  d^3 p_2 \, \int  d^3p_3 \, \int  d \Omega_4 \, v_{12}\, 
\frac{d\sigma_{12 \rightarrow 34}}{d\Omega}\,  
\delta^3 \left(\vec{p}_1 + \vec{p}_2 
- \vec{p}_3 - \vec{p}_1 \right) \nonumber \\
&\times& (f_3(\vec{r},\,\vec{p}_3,\,t)\, 
f_4(\vec{r},\,\vec{p}_4,\,t)\, \bar{f}_1(\vec{r},\,\vec{p}_1,\,t)\,
\bar{f}_2(\vec{r},\,\vec{p}_2,\,t)) \nonumber \\ &-&  \, \, \, 
f_1(\vec{r},\,\vec{p}_1,\,t)\, 
f_2(\vec{r},\,\vec{p}_2,\,t)\, \bar{f}_3(\vec{r},\,\vec{p}_3,\,t)\,
\bar{f}_4(\vec{r},\,\vec{p}_4,\,t)), 
\eea
where $\frac{d\sigma_{12 \rightarrow 34}}{d\Omega}$ is the in-medium differential nucleon-nucleon 
cross section, $\bar{f}_i = 1 - f_i\, (i = 1, .., 4)$ the Pauli-blocking
factors and $v_{12}$ the relative velocity between the nucleons 
$N_1$ and $N_2$ in their center-of-mass system. The factor $g = 2$ in 
(\ref{collint}) stands for the 
spin degeneracy of the nucleons whereas $\sum_{2,3,4}$ stands for the sum over
the isospin degrees of freedom of particles $N_2, N_3$ and $N_4$.
\\
\\
For energies above the pion-production threshold one also has to account 
for inelastic processes such as direct meson production channels
or the excitation/deexcitation  
of higher baryon resonances. In the CBUU-model - to be described here - 
we explicitly propagate the
mesonic degrees of freedom $\pi$, $\eta$, $\rho$ and a scalar meson $\sigma$
that simulates correlated $2\pi$ pairs in the isospin $0$-channel. 
Besides the nucleon and 
the $\Delta(1232)$ we, furthermore, include all baryonic resonances up to 
a mass of $1950$ MeV/c$^2$: i.e. $N(1440)$, $N(1520)$, $N(1535)$, $\Delta(1600)$, 
$\Delta(1620)$, $N(1650)$, $\Delta(1675)$, $N(1680)$, $\Delta(1700)$,
$N(1720)$, $\Delta(1905)$, $\Delta(1910)$ and $\Delta(1950)$, 
where the resonance properties are adopted from the PDG \cite{pdg}. \
Introducing one-body phase-space distribution functions for each particle 
type leads to equations similar to eq. (\ref{buueq}) for each hadron. 
Since the different particle species
mutually interact the integro-differential
equations are coupled by the collision integrals. Schematically  
one can write down the set of coupled equations in the 
following way: 
\bea
D f_N & = & I_{coll}[f_N, f_{\Delta(1232)}, ..., f_{\Delta(1950)}, 
f_\pi, f_\rho, f_\eta, f_\sigma] \nonumber \\
D f_{\Delta(1232)} & = & I_{coll}[f_N, f_{\Delta(1232)}, ..., f_{\Delta(1950)}, 
f_\pi] \nonumber \\
...                 & = &  ... \nonumber \\
D f_{N(1535)} & = & I_{coll}[f_N, f_{\Delta(1232)}, ..., f_{\Delta(1950)}, 
f_\pi, f_\rho, f_\eta, f_\sigma] \nonumber \\
D f_{\Delta(1600)} & = & I_{coll}[f_N, f_{\Delta(1232)}, ..., f_{\Delta(1950)}, 
f_\pi, f_\rho,f_\sigma] \nonumber \\
..                 & = &  .. \nonumber \\
D f_{\Delta(1950)} & = & I_{coll}[f_N, f_{\Delta(1232)}, ..., f_{\Delta(1950)}, 
f_\pi, f_\rho, f_\sigma] \nonumber \\
D f_{\pi} & = & I_{coll}[f_N, f_{\Delta(1232)}, ..., f_{\Delta(1950)}, 
f_\pi, f_\rho, f_\sigma] \nonumber \\      
D f_{\eta} & = & I_{coll}[f_N, f_{N(1535)}] \nonumber \\
D f_{\rho} & = & I_{coll}[f_N, f_{N(1440)}, ..., f_{\Delta(1950)}, 
f_\pi] \\
D f_{\sigma} & = & I_{coll}[f_N, f_{N(1440)}, ..., f_{\Delta(1950)}, 
f_\pi] 
\label{coupledeq},      
\eea
where $Df$ abbreviates the left side of the Vlasov-equation. The collision
integrals on the r.h.s. of eq. (\ref{coupledeq}) have formally the same
structure as the one given in eq. (\ref{collint}).\ 
In addition to elastic 
scattering processes they now also contain all allowed transition rates.
Denoting the nucleon by $N$ and the baryon resonances listed above by $R$ and 
$R'$, we include explicitly the following channels:
\begin{itemize}
\item{elastic baryon-baryon collisions}
\bea
N N & \leftrightarrow & N N \nonumber \\
N R & \leftrightarrow & N R \non
\eea
The elastic $NN$-cross section is described within the parametrization 
by Cugnon  \cite{bertsch88,cugnon82}, while the cross section for 
elastic $NR$-scattering
is evaluated by using an invariant matrix element which is extracted 
from the $NN$-cross section (cf. Section 3). 
For $NR$-scattering we also allow for a change in the resonance
mass according to the corresponding distribution function (cf. Section 3).
\item{inelastic baryon-baryon collisions}
\bea
N N & \leftrightarrow & N R \non \\
N R & \leftrightarrow & N R'\non \\
N N & \leftrightarrow & \Delta(1232) \Delta(1232) \non
\eea
For the cross section of the reaction $NN \rightarrow N\Delta(1232)$ we use
the result of the OBE-model calculation by Dimitriev and
Sushkov \cite{dimitriev86}. In order to
obtain the $NN\leftrightarrow NR$-cross sections for any of the higher resonances
we exploit the resonance model described in more detail in Section 3,
while the parametrization for the
$NN\leftrightarrow\Delta(1232)\Delta(1232) $-cross section is adopted from 
Huber and Aichelin \cite{aichelin94}.
\item{inelastic baryon-meson collisions}
\bea
R & \leftrightarrow & N \pi \non \\
R & \leftrightarrow & N \pi \pi \non \\
  & \leftrightarrow & \Delta(1232) \pi,\, N(1440)\pi, \, N \rho,\, N \sigma \non \\
N(1535) & \leftrightarrow & N \eta \non \\
N N & \leftrightarrow & N N \pi \non
\eea
Besides the production and absorption in baryon-baryon collisions,  
baryonic resonances can also be populated in baryon-meson collisions and 
subsequently decay to baryons and mesons again. 
A detailed discussion of the respective cross
sections and decay widths is given in Section 3. Here we
note that the  $2\pi$-decay of higher resonances is modelled via
subsequent two-body decays as indicated above.
\item{meson-meson collisions}
\bea
\rho & \leftrightarrow & \pi \pi \, \,\, \, \,  \mbox{(p-wave)}\non \\
\sigma & \leftrightarrow & \pi \pi \, \, \, \, \, \mbox{(s-wave)} \non
\eea
For the pure mesonic cross sections we use
the Breit-Wigner parametrizations given in Section 3.
\end{itemize}

\end{subsection}
\begin{subsection}{The Test-Particle Method} \label{sec_tpm}
The CBUU-equations (\ref{coupledeq}) are solved by means of the test-particle
method, where the phase-space
distribution function \fn (e.g. for nucleons) is represented by a sum over 
$\delta$-functions: 
\be
f( \vec{r},\, \vec{p},\, t) =\frac{1}{N}\, \sum_{i=1}^{N \times A}
\delta \left(\vec{r}- \vec{r}_i(t)\right) \times \delta \left( \vec{p}
- \vec{p}_i(t) \right).
\label{tpansatz}
\ee
Here $N$ denotes the number of test-particles per nucleon while $A$ is
the total number of nucleons participating in the reaction. Inserting the 
ansatz (\ref{tpansatz}) into the CBUU-equations (\ref{coupledeq}) leads
to the following equations of motion for the test-particles: 
\bea
\frac{d \vec{r}_i(t)}{d t} & = & \frac{\vec{p}}{E} + \frac{m^*}{E}\;
\vec{\nabla}_p \; U(\vec{r}_i,\, \vec{p}_i(t) ) \non \\
\frac{d \vec{p}_i(t)}{d t} & = & - \frac{m^*}{E}\;
\vec{\nabla}_r \; U(\vec{r}_i,\, \vec{p}_i(t) ). \label{tpeoms}
\eea
Again these equations of motion are consistent with those from
a covariant transport equation \cite{kweber93} when using only 
scalar self-energies for the baryons. Thus the solution of the CBUU-equations within the test-particle method reduces to the
time evolution of a system of classical point particles according to eq. (\ref{tpeoms}).
For the actual numerical simulation we discretize the time $t$ and integrate the
equations of motion employing a predictor-corrector
method \cite{stoer80}. We want to note that in our model pions are treated 
as 'free' particles, except for the Coulomb interaction. 
It remains to be seen if pion selfenergies as suggested in refs. 
\cite{ehehalt,koo} will alter our results. 
\end{subsection}
\begin{subsection}{The Collision Integrals}\label{sec_evcolint}
The collision integrals occurring in eq. (\ref{coupledeq}) contain either
particle-particle collisions or the decay of baryonic or mesonic resonances.
\\
For particle-particle collisions we employ the following prescription: 
The test-particles collide with each other as in conventional cascade
simulations with reaction probabilities that are calculated on the basis of free
cross sections. The numerical implementation additionally accounts for
the Pauli-blocking of the final states while the collision
sequence is calculated within the Kodama algorithm \cite{kodama84} (cf. ref. 
\cite{wolf90}) which is an approximately covariant prescription.\\
The decay of a resonance, furthermore, is determined by its width $\Gamma(M)$.
However, during the course of a heavy-ion collision the resonance may also
decay due to collisions with other particles (eq. $NR \rightarrow NN$).
This collisional broadening of a resonance is described by the collision
integral for particle-particle collisions. Here we will concentrate 
on the numerical
method used to account for the first decay mechanism. All resonances treated
in the CBUU-model are allowed to decay into a two-particle final state, i.e. 
in every timestep of
the simulation we calculate the decay probability $P$ for each resonance 
assuming an exponential decay law 
\be
P = 1 - e^{-\Gamma(M)/\hbar \gamma \Delta t}, \label{resdecprob}
\ee
where $\Delta t$ is the time step size of our calculation, $\Gamma(M)$ is the
energy-dependent width of the resonance and $\gamma$ the Lorentz factor related
to the velocity of the resonance with respect to the calculational frame. 
We then decide by means of a Monte Carlo
algorithm if the resonance may decay in the actual timestep 
and to which final state it may go.
If the chosen final state contains a nucleon, which e.g. 
is Pauli-blocked, we reject the resonance decay.
\end{subsection}
\end{section}
\begin{section}{Explicit numerical Implementations}
\begin{subsection}{Nuclear Mean-Field Potentials}\label{mfpintro}
Particles propagating inside nuclear matter are exposed to the mean-field
potential generated by all other particles. From Dirac-phenomenological
optical-model calculations \cite{kweber93,arnold79} it is known that
elastic nucleon-nucleus scattering data can only be described when using
proper momentum-dependent potentials. Here we employ the 
momentum-dependent mean-field potential proposed by Welke et al. 
\cite{welke88}, i.e.
\be
U^{nr}(\vec{r}, \, \vec{p}) = A \frac{\rho}{\rho_0} +
B \frac{\rho}{\rho_0}^\tau + 2 \frac{C}{\rho_0} \int d^3 p'\frac{f(\vec{r},\vec{p'})}
{1 + \left(\frac{\vec{p}-\vec{p'}}{\Lambda}\right)^2}.
\label{welkepot}
\ee
As an extension of the momentum-independent Skyrme type 
potentials for nuclear matter \cite{cassing90,welke88} the parametrization
(\ref{welkepot}) has no manifest Lorentz-properties.
However, definite Lorentz-properties are required for a transport model at 
relativistic energies. To achieve this goal
we evaluate the non-relativistic mean-field potential $U^{nr}$ in the
local rest frame (LRF) of nuclear matter which is defined by the frame
of reference with vanishing local vector baryon current ($\vec{j}(r,t) = \vec{0}$).
Discarding vector potentials in the LRF we then equate the 
expressions for the single-particle energies using the
non-relativistic potential $U^{nr}$ and the scalar potential $U$ by 
\be
\sqrt{p^2 + m^2} + U^{nr}(\vec{r},\,\vec{p}) =  \sqrt{p^2 + \left(
m + U \left( \vec{r},\,\vec{p} \right) \right)^2}.
\label{defuscal}
\ee
Eq. (\ref{defuscal}) now allows to extract the scalar mean-field potential 
$U(\vec{r},\,\vec{p})$ which we will use throughout our
calculations for the baryons. Due to the relativistic dispersion 
relation for the quasiparticle (\ref{defuscal}) the scalar
potential $U(\vec{r},\,\vec{p})$  now has 
definite Lorentz-properties contrary to $U^{nr}$.
This enables us to guarantee energy conservation in each two-body collision
and in resonance decays (eq. $N_1+N_2 \rightarrow N_3+N_4$) as 
\bea
E_1 + E_2 & = & E_3 + E_4 \non \\
\sqrt{\vec{p}_1^2 + m_1^{*\, 2}} +\sqrt{\vec{p}_2^2 + m_2^{*\, 2}}
& = & \sqrt{\vec{p}_3^2 + m_3^{*\, 2}} + \sqrt{\vec{p}_4^2 + m_4^{*\, 2}}.
\label{defencons}
\eea
Since the collision integrals (\ref{coupledeq}) are evaluated in the 
center-of-mass system of the colliding particles or in the restframe of the 
decaying
resonance, while the testparticle equations of motion are generally integrated in
the center-of-mass system of the heavy-ion reaction, energy conservation is not a priori
fulfilled when employing non-relativistic potentials without definite properties
under Lorentz transformations. \\
For our calculations we will use a 
(momentum-dependent) equation of state (EOS)
for nuclear matter with an incompressibility of K = 250 MeV ( A = -29.253 MeV,
B = 57.248 MeV, C = -63.516 MeV, $\tau$ = 1.760, $\Lambda$ = 2.13 1/fm.) \
The actual calculation of the mean-field potential according to eq. (\ref{welkepot}), however,  
is too involved for practical purposes since one has to perform double 
integrations for the momentum-dependent part of $U^{nr}(\vec{r},\ \vec{p})$. 
We thus determine the potentials within the Local Thomas-Fermi approximation,  
where the integral over the phase-space distribution function
in (\ref{welkepot}) can be performed analytically \cite{welke88}.
With
\bea
f(\vec{r}, \vec{p}) = \frac{4}{(2\pi\hbar)^3} \, \Theta(p_F -p)\non,
\eea
where the factor $4$ stems from the sumation over spin and isospin,
the integral for the momentum-dependent part of the potential $U^{nr}$ reads:
\bea
 \int d^3 p'\frac{f(\vec{r},\vec{p'})}
{1 + \left(\frac{\vec{p}-\vec{p'}}{\Lambda}\right)^2} = 
& & \frac{4}{(2\pi\hbar)^3} \,\pi \Lambda^3 \left[ 
\frac{p_F^2 + \Lambda^2 - p^2}{2p\Lambda}
ln\frac{\left(p + p_F\right)^2 + \Lambda^2}{\left(p - p_F\right)^2 + \Lambda^2}
\right.  \label{welkepottfa}    \\ & & \left. + \frac{2p_F}{\Lambda} - 2 \left[arctan\frac{p+p_F}{\Lambda}
- arctan\frac{p-p_F}{\Lambda} \right] \right].   \non
\eea
We employ a smeared baryon density \cite{cassing90} when evaluating expression (\ref{welkepottfa})
rather than that obtained directly from the test-particle distribution in order 
to avoid unphysical statistical fluctuations in the density and in the 
mean-field potentials. 
\end{subsection}
\begin{subsection}{The Coulomb Potential}\label{sec_coul}
Charged baryons and mesons are additionally exposed to the Coulomb potential $V_c(\vec{r})$
generated by all charged particles. Since in our present approach mesons 
are propagated as free particles with respect to the nuclear mean-field, 
the Coulomb force
\be
\vec{F}_c \left(\vec{r}\right) = - q \,
\vec{\nabla}_r V_c \left(\vec{r}\right)
\label{coul_force}
\ee
is the only force acting on a meson with charge $q$. For charged baryons the force  
$F_c$ (\ref{coul_force}) represents an additional term in the equations of motion (\ref{tpeoms}).
The Coulomb potential $V_c\left(\vec{r}\right)$ is obtained by solving the Poisson-equation by means
of the {\it Alternating-Direction Implicit Iterative} 
(ADI-)algorithm \cite{varga62}.
\end{subsection}
\begin{subsection}{Resonance Properties and Decay Widths}\label{sec_resprop}
Within the CBUU-approach the resonances are treated as "on-shell" 
particles with
respect to their propagation and the evaluation of cross sections. To account for
their "off-shell" behaviour, we 
distribute the resonance masses according to a Lorentzian 
distribution function (see
sec. (\ref{sec_inelnn})), which is determined by the
mean resonance masses $M_R$ and the total and partial decay 
widths $\Gamma_R$ at mass
$M_R$. For the baryonic resonances these explicit parameters are given in
table \ref{tab_baryprop}. For the $\rho$-meson we use  a mean mass
$M_R = 770$ MeV/c$^2$ and for the decay width at resonance $\Gamma_R = 118$
MeV. The corresponding values for the $\sigma$-meson are $M_R = 800$ MeV/c$^2$
and $\Gamma_R = 800$ MeV. \\
\begin{table}[t]
\begin{center}
\begin{tabular}{|c||c||c||c|c|c|c|c|c|} 
\hline
& & &\multicolumn{6}{|c|}{branching ratio [\%]}\\
\cline{4-9}
resonance & $\overline{\left| \mbox{$\cal M$}^2 \right|}/ 16 \pi$ & $\Gamma_R$ & & & 
\multicolumn{4}{|c|}{N $\pi \pi$} \\
\cline{6-9}
& [$mb\,GeV^2$]& [$MeV$] & 
\raisebox{1.5ex}[-1.5ex]{N $\pi$} & 
\raisebox{1.5ex}[-1.5ex]{N $\eta$} &
$\Delta \pi$ & N $\rho$ &
N $(\pi \pi)_{s-wave}^{I=0}$&
$N(1440) \pi$\\ 
\hline
$\Delta$(1232) & - & 120 & 100 &    0 &     0 &  0 &  0 & 0\\
\hline
$N$(1440) &14& 350  & 65    &  0 &  25 &    0 & 10 & 0  \\
\hline
$N$(1520) &4& 120 & 55 &  0 &     25 &  15 & 5 & 0\\
\hline
$N$(1535) &8, 40& 203 &50     &  45 &  0 &    2 & 0 &  3  \\
\hline
$\Delta$(1600) &68& 350 &15 &  0 & 75 &    0 & 0 & 10  \\
\hline
$\Delta$(1620) &68& 150 & 30 &     0 &  60 &    10 & 0 & 0 \\
\hline
$N$(1650) &4& 150 & 80 &  0 & 7 & 5 &  4 & 4  \\
\hline
$\Delta$(1675) &68& 150 & 45 &     0 &     55 &  0 &  0 & 0 \\
\hline
$N$(1680) &4& 130 & 70 &  0 &     10 &5 & 15&0  \\
\hline
$\Delta$(1700) &7& 300 & 15 &     0 & 55 &30 &  0 & 0 \\
\hline
$N$(1720) &4& 150 & 20    &  0 &  0&   80 & 0 & 0 \\
\hline
$\Delta$(1905) &7& 350 & 15       &  0 &  25 & 60 & 0 & 0 \\
\hline
$\Delta$(1910) &68& 250 & 50             &  0 & 50 & 0 & 0 & 0  \\
\hline
$\Delta$(1950) &14& 300 & 75       & 0 & 25 & 0 & 0 & 0 \\
\hline
\end{tabular}
\end{center}
\caption{Decay widths and decay channels for the baryonic resonances.
The data for these channels are taken from \protect\cite{pdg} 
and \protect\cite{krusche95} in case of the $N(1535)$. The second column
contains the averaged matrix-elements for the production of the 
baryon resonances in nucleon-nucleon collisions (cf. sec. \protect\ref{sig_nnnr}).}
\label{tab_baryprop}
\end{table}
In the following we list the parametrizations used for the decay widths $\Gamma(M)$.
\begin{itemize}
\item{$1\pi$-decay width for the $\Delta(1232)$}\\
For the $\Delta(1232)$-decay we adopt the parametrization given by Koch
et al. \cite{moniz84}
\be
\label{moniz}
\Gamma_{Moniz}(q)=\Gamma_R \frac{M_{\Delta}}{M} \left( \frac{q}{q_r} \right)^{3}
\left( \frac{q_r^2+\delta^2}{q^2+\delta^2} \right)^{2},
\ee
where $M$ is the actual mass of the $\Delta(1232)$ and
$M_\Delta = 1232$ MeV/c$^2$. $q$ and $q_r$ are the pion three-momenta
in the restframe of the resonance with mass $M$ and $M_\Delta$, 
respectively. The parameter $\delta$ in the cutoff function has a value
 $\delta = 0.3$ GeV/c$^2$.
\item{$1\pi$-decay width for the higher baryon resonances} \\
The $1\pi/\eta$-decay widths for the higher baryon resonances are
given by
\be
\label{standardbreite}
\Gamma(q)=\Gamma_R \left( \frac{q}{q_r} \right)^{2l+1}
\left( \frac{q_r^2+\delta^2}{q^2+\delta^2} \right)^{l+1},
\ee
where $l$ is the angular momentum of the emitted pion or $\eta$ and
$q$ and $q_r$ are the momenta of the pion or $\eta$ in the restframe
of the decaying resonance as defined above. In this case, we use
\be
\delta^2=(M_R-M_N-m_{\pi})^2+\frac{\Gamma_R^2}{4}
\ee
\item{$2\pi$-decay width for baryon resonances} \\
The $2\pi$-decay of the higher baryon resonances is described by a
two-step process. First, a higher lying baryonic resonance decays into a $\Delta(1232)$ or
$N(1440)$ and a pion or into a nucleon and $\rho$- or $\sigma$-meson.
The new resonances then propagate through the nuclear medium and
eventually decay into a nucleon and pion or a nucleon and two pions
\be
R \rightarrow r \, b \rightarrow N \, \pi \, \pi. \label{twopitwostep}
\ee
Here R denotes the higher baryonic resonances, r stands for a $\Delta(1232)$,
 $N(1440)$,  $\rho$ or  $\sigma$ and b is a nucleon or 
pion, respectively. Due to the fact that a further resonance appears 
in the first step
of reaction (\ref{twopitwostep}) one has to integrate the corresponding
distribution function over the mass $\mu$ of the intermediate resonance
$r$ in order to obtain the $2\pi$-decay width
\bea
\label{zweipionenbreite}
\Gamma_{R \to r\,b}(M)&=& \frac{P_{2\pi}}{M} \int_0^{M-m_b} d\mu \, p_f \,\frac{2}{\pi}
\frac{\mu^2\, \Gamma_{r,tot}(\mu)}{(\mu^2-m_r^2)^2+\mu^2\,\Gamma_{r,tot}^2(\mu)} 
\nonumber
\\
&&
\times \left( \frac{\left(M_R-M_N-2m_{\pi}\right)^2+\delta^2}
{\left(M-M_N-2m_{\pi}\right)^2+\delta^2}\right)^2,
\eea
where $P_{2\pi}$ is the branching ratio for the $2\pi$-decay of the baryonic resonance
$R$ and $p_f$ denotes the momentum of $r$ and $b$ in the restframe of $R$. Since
the integral in eq. (\ref{zweipionenbreite}) for high resonance masses $M$
is proportional to $M$, we introduce a cutoff function to avoid $\Gamma_{R \to r\,a}(M)$
to diverge for high masses. The parameter $\delta$ used is 0.3 GeV.
\item{decay width for meson resonances} \\
The decay width of the meson resonances is parametrized similarly to
that of the $\Delta(1232)$,
\be
\Gamma(M) = \Gamma_r \frac{M_r}{M}
\left( \frac{q}{q_r} \right)^{2J_r+1} 
\frac{q_r^2+\delta^2}{q^2+\delta^2}, \label{mesdecaywidth}
\ee
where $M_r$ and $M$ are the mean mass and the actual mass of the 
meson resonance.
$q$ and $q_r$ are defined as in eq. (\ref{moniz}) while $J_r$ 
is the spin of the
resonance and $\Gamma_r$ the decay width for a resonance with mass $M_r$.
For the parameter $\delta$ in the cutoff function we use again $\delta $ = 
0.3 GeV.
\end{itemize}
\end{subsection}
\begin{subsection}{Meson-Baryon Cross Sections}\label{sec_xmesbar}
In order to describe meson-baryon scattering in the framework of our
resonance picture we use a Breit-Wigner formulation for the cross sections
(eq. $\pi N \rightarrow \pi N$)
\begin{equation}
\label{breit}
\sigma_{ab \rightarrow R \rightarrow cd} = 
\frac{2J_R+1}{(2S_a+1)(2S_b+1)}\, \frac{4 \pi}{p_i^2} 
\frac{s\,\Gamma_{R \rightarrow ab}\, \Gamma_{R \rightarrow cd}}
{(s-M_R^2)^2+s\, \Gamma_{tot}^2} .
\end{equation}
In eq. (\ref{breit}) $ab$ and $cd$ denote the baryon and the meson in the
final and initial state of the reaction and $R$ is the intermediate baryon
resonance. $J_R$, $S_a$ and $S_b$ are the spins of the baryon resonance and
the particles in the initial state of the reaction. \\
The solid line in fig. 
\ref{fig_pinabstot} shows the total $\pi^--p$-cross section within our
model in comparison to the experimental data from \cite{landolt}. 
To calculate this
cross section we replace the partial widths $\Gamma_{R \to cd}$ in
eq. (\ref{breit}) by the total widths of the baryonic resonances and sum up the
contributions from all resonances. The dashed, the dotted and the dash-dotted lines
in fig. \ref{fig_pinabstot} show the contributions from the $\Delta(1232)$,
the $N(1440)$ and the $N(1535)$ separately.
\begin{figure}
\vspace{-2.0
cm}
\begin{center}
\hspace{-35mm}{\epsfig{file=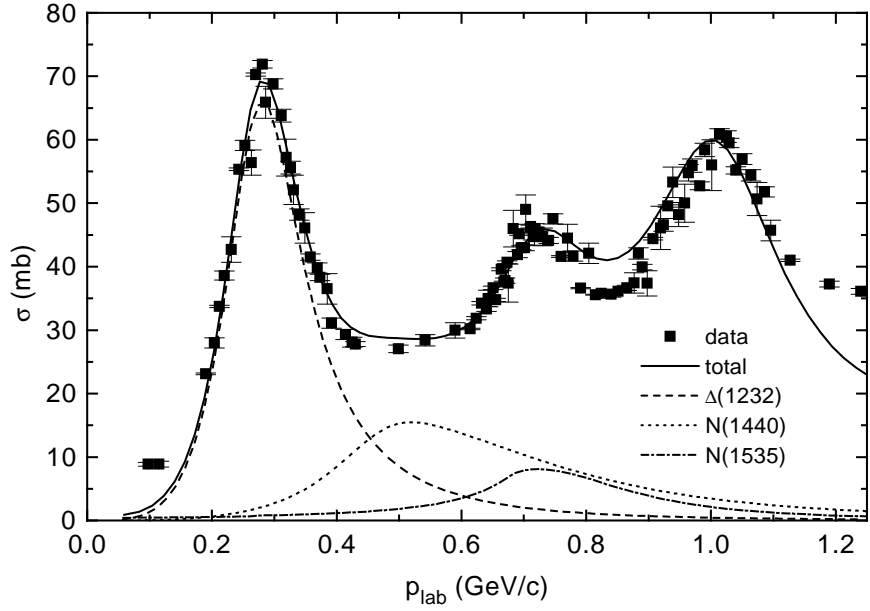,width=90mm}}
\vspace{-4.5cm}
\end{center}
\caption{The calculated total $\pi^--p$-cross section in comparison to the
data from \protect\cite{landolt} is given by the solid line.
The dashed, the dotted and the dashed-dotted lines show the contributions
from the $\Delta(1232)$, $N(1440)$ and $N(1535)$, respectively.}
\label{fig_pinabstot} 
\end{figure}
\begin{figure}
\vspace{-2.0cm}
\begin{center}
\hspace{-35mm}{\epsfig{file=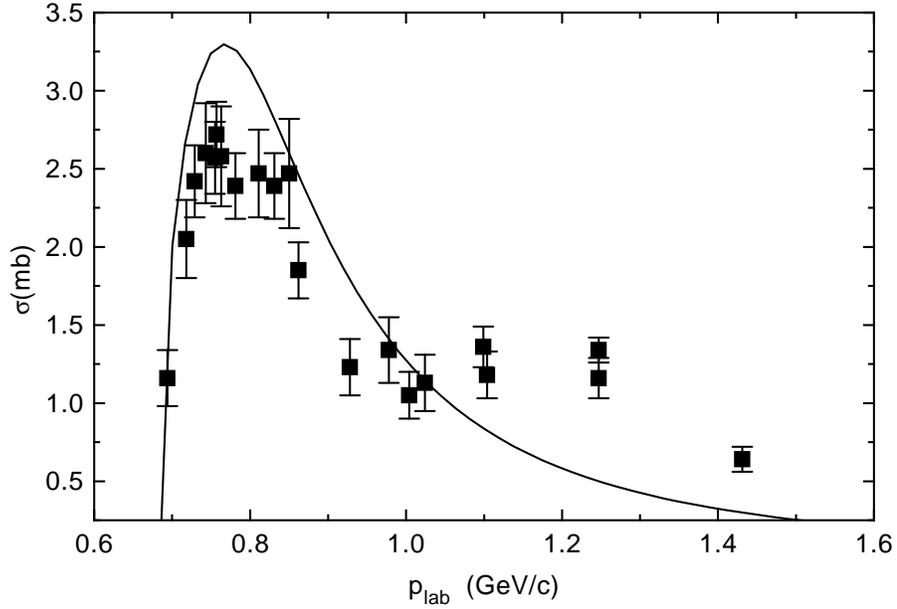,width=90mm}}
\vspace{-4.5cm}
\end{center}
\caption{The solid line shows the cross section for the reaction
$\pi^- p \rightarrow \eta n$ as obtained from eq. (\protect\ref{breit}).
The data are taken from \protect\cite{landolt}.}
\label{fig_etaabs}  
\end{figure}
In fig. \ref{fig_etaabs} we display the resulting cross section for the
 $\pi^- p \rightarrow \eta n$ reaction. Here only the $N(1535)$ contributes
since this is the only baryonic resonance that couples to the $\eta$ 
within our model space (cf. table \ref{tab_baryprop}). Both cross sections 
are obviously well described up to $p_\pi \approx$ 1.0 GeV.\\
Eq. (\ref{breit}) is also used to determine the cross sections for
$\sigma$- and $\rho$-production in $\pi-\pi$-collisions weighting
with the corresponding
spins of the mesonic resonances and the pions in the initial state of the
reaction.
\end{subsection}
\begin{subsection}{Inelastic Baryon-Baryon Cross Sections}\label{sec_inelnn}
In this section we describe the concepts and parametrizations used to
implement the cross sections for resonance, pion and $\eta$ production in the
CBUU-model.
\begin{subsubsection}{The $NN \to N\Delta(1232)$ Cross Section}\label{sig_nnnd}
For the $NN \to N \Delta(1232)$ cross section we employ the result of the
OBE-model calculation by Dimitriev and Sushkov \cite{dimitriev86} using 
$u$- and $t$-channel Born-diagrams. The parameters
of the model ($NN\pi$-, $N\Delta(1232)\pi$-coupling constant and a parameter
in the $\pi$-formfactor) are chosen to reproduce the experimental $pp \to N\Delta^{++}$
cross section. For the implementation of this cross section into the CBUU-model
we replace the parametrization for the $ \Delta(1232)$-width given in \cite{dimitriev86}
by the Moniz parametrization (\ref{moniz}). An example for the resulting
mass and angular differential cross sections for an invariant
energy of $2.31$ GeV is given in fig. \ref{fig_dimi} in
comparison to the experimental data \cite{dimitriev86}. The cross sections
for the other isospin channels follow from 
the $pp\to n\Delta^{++}$ cross section by applying isospin symmetry.
\begin{figure}
\vspace{-2cm}
\begin{center}
\hspace{0mm}{\epsfig{file=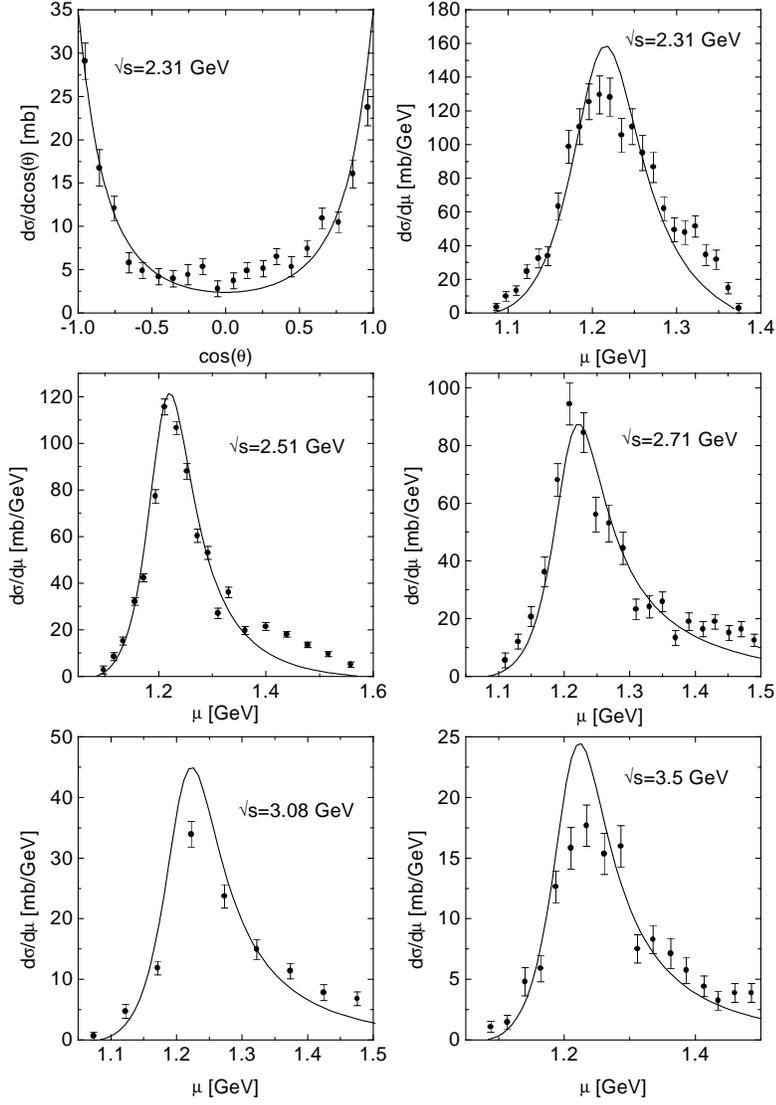,width=112mm}}
\vspace{-1cm}
\end{center}
\caption{Comparison of the $pp \to n \Delta(1232)$ angular and mass
differential cross section at an invariant energy of $2.31$ GeV (solid lines)
with the experimental data \protect\cite{dimitriev86} (filled circles).}
\label{fig_dimi}
\end{figure}
\end{subsubsection}
\begin{subsubsection}{The $NN \to NR$ Cross Section}\label{sig_nnnr}
In order to obtain the production cross sections for the higher
baryon resonances in nucelon-nucelon collisions we fit the corresponding
matrix-elements to available data for $\eta$-, $1\pi$-, $\rho$- and
$2\pi$-production in nucleon-nucleon reactions. Therefore we assume that the
$\eta$-, $1\pi$-, $\rho$- and
$2\pi$-production in nucleon-nucleon collisions above the $\Delta(1232)$
excitation proceed only through
these resonances via subsequent two-step processes. The $\eta$, single pion
and $\rho$ production are described by the creation of a baryonic resonance in
a nucleon-nucelon collision and its subsequent decay into $\eta$, pion or $\rho$
\be
\label{twostepnnnr}
NN \to NR \to NN \pi/\eta/\rho.
\ee
We assume that the $2\pi$-production in nucleon-nucleon collisions proceeds
either through the excitation of two $\Delta(1232)$ and their subsequent
decay into a nucleon and a pion or through the excitation of higher lying
baryonic resonances and their subsequent decay into a nucleon and two
pions (see section (\ref{sec_resprop})).
\\
\begin{itemize}
\item{The general expression for the cross section}\\
For the derivation of the cross section we assume that in a collision of two
nucleons $a+b \to R + c$ a baryonic resonance $R$ and a nucleon $c$
are produced. Then resonance $R$ decays into a two-body final 
state$R \to de$. Assuming spinless particles
the invariant matrix element is given by
\be
\label{matrixele1}
\mbox{$\cal M$}_{a b \rightarrow c de} =
\mbox{$\cal M$}_{a b \rightarrow Re}\, P_R \,
\mbox{$\cal M$}_{R \rightarrow c d},
\ee
where $P_R$ is the propagator of the intermediate baryonic resonance $R$ and
${\cal M}_{a b \rightarrow Re}$ and ${\cal M}_{R \rightarrow c d}$ are the
matrix elements for the reactions $ a+b \to R +e$ and $R\to c+d$, respectively.
We start from the general expression for the cross section
\bea
d\sigma_{ab \rightarrow cde}  & = &
\frac{(2 \pi)^4}{4\, p_i\, \sqrt{s}}\, \delta^4 
(p_a+p_b-p_c-p_d-p_e) \overline{\left| \mbox{$\cal M$}_{a b \rightarrow c de} \right|^2}
\nonumber
\\ & &
\times\frac{d^3 p_c}{(2 \pi)^3\, 2 E_c}
\frac{d^3 p_d}{(2 \pi)^3\, 2 E_d}
\frac{d^3 p_e}{(2 \pi)^3\, 2 E_e},
\label{crossgen}
\eea
where $\sqrt{s}$ is the invariant energy of the particles in the initial state and
$p_i$ is their CMS momentum.
$\overline{\left| \mbox{$\cal M$}_{a b \rightarrow c de}\right|^2}$
is the square of the invariant matrix element averaged over the spin of particles
$a$ and $b$ and summed over the spins of the particles in 
the final state of the reaction. Assuming that 
the square of the matrix element factorizes as
\be
\label{matrixfactori}
\overline{\left| \mbox{$\cal M$}_{ab \rightarrow cde} \right| ^2} =
\overline{\left| \mbox{$\cal M$}_{ab \rightarrow Re} \right| ^2}
\left| P_R \right|^2
\overline{\left| \mbox{$\cal M$}_{R \rightarrow cd} \right| ^2},
\ee
we obtain for the cross section as a function of the resonance mass $\mu$ 
\bea
\label{fincrossexpr}
\frac{d\sigma_{ab \to Re \to cde}}{d\mu} =
\sigma_{ab \to Re}(\mu)\, \frac{2}{\pi}\,
\frac{\mu^2 \,\Gamma_{R \to cd}}{(\mu^2-M_R^2)^2+\mu^2\, \Gamma_{tot}^2}.
\eea
Here, $\sigma_{ab \to Re}(\mu)$ is the cross section for producing a baryonic
resonance $R$ with a fixed mass $\mu$
\bea
\label{singlecross}
\sigma_{ab \to Re}(\mu)=\frac{1}{64\, \pi^2\, s\, p_i} \int d\Omega\,
p_f
\overline{\left| \mbox{$\cal M$}_{ab->Re}(\mu) \right|^2},
\eea
with $p_f$ and $p_i$ denoting the CMS momenta of the particles in the  
final and initial state of the reaction $a+b \to R+e$, respectively, while 
$s$ is the squared invariant energy of this reaction. When 
evaluating the production cross sections for the higher baryonic resonances 
we assume 
$\overline{\left| \mbox{$\cal M$}_{ab->Re}(\mu) \right|^2}$ to be constant 
for all baryonic resonances except for the $\Delta(1232)$ (cf. section
(\ref{sig_nnnd})).\\
\item{$\eta$-production cross section} \\
Since in our model only the $N(1535)$ couples to the $\eta$-meson,
we obtain the cross section for $\eta$-production in nucleon-nucleon 
collisions from eq. (\ref{fincrossexpr}) by using the $N(1535)$ as the 
intermediate baryonic resonance and regarding $NN\eta$ as the final 
state $cde$. The unknown squares of the matrix elements 
$ \overline{\left| \mbox{$\cal M$}_{ab->Re} \right|^2}$ are obtained by 
fitting the available experimental data for $\eta$-production in 
nucleon-nucleon collisions. The matrix element for $N(1535)$ 
production in proton-proton collisions then gives 
\be
\overline{\left|\mbox{$\cal M$}_{p\,p\to p\,N^+(1535)} \right|^2}
= 16\pi\times 8 {\rm \,mb \, GeV^2}. 
\label{matrixpos1}
\ee
The result of this fit for the reaction $pp \rightarrow pp \eta$ 
is displayed in fig. \ref{fig_matrixpos1} in comparison
to the data \cite{chiavasa94,landolt}. 
\begin{figure}
\vspace{-2cm}
\begin{center}
\hspace{-35mm}{\epsfig{file=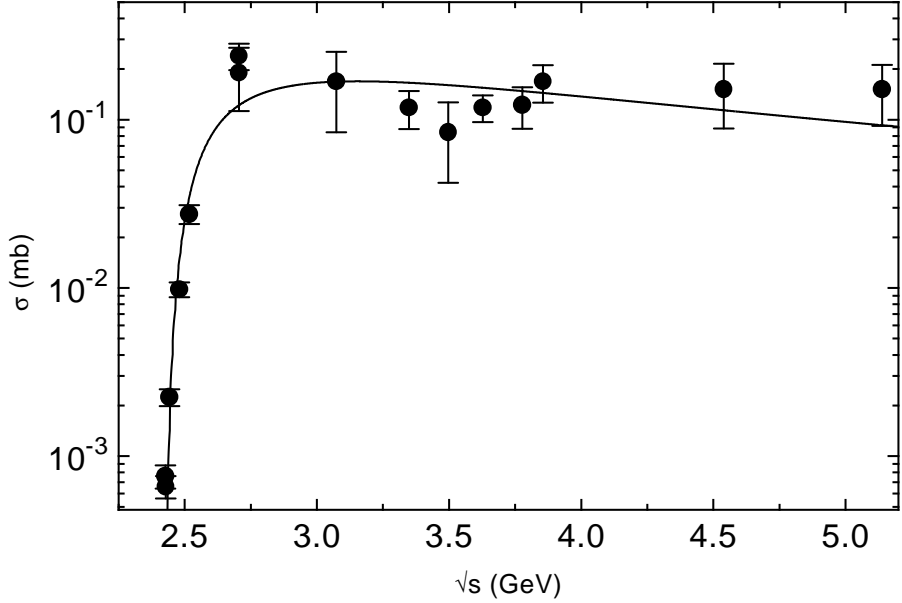,width=90mm}}
\vspace{-4.5cm}
\end{center}
\caption{Cross section for the reaction $pp \to pp\eta$ obtained from the 
resonance model using the matrix element given in eq. 
(\protect\ref{matrixpos1}) (solid line) in comparison to the experimental data
\protect\cite{chiavasa94,landolt}.}
\label{fig_matrixpos1}
\end{figure}
It is known from \cite{chiavasa94} that the cross section for $\eta$-production
in proton-neutron collisions is about a factor $5$ larger than that for 
proton-proton collisions. Consequently we use 
\be
\overline{\left|\mbox{$\cal M$}_{p\,n\to p\,N(1535)} \right|^2} 
=\overline{\left|\mbox{$\cal M$}_{p\,n\to n\,N^+(1535)} \right|^2}  
= 16\pi\times 40 {\rm \,mb \, GeV^2}. 
\label{matrixpos}
\ee
\item{$1\pi$-production cross section}\\
In order to evaluate the $1\pi$-production cross sections in nucleon-nucleon 
collisions we sum up the contributions from all resonances contributing to 
a specific channel incoherently. For $\Delta(1232)$-production we use 
the cross section given in sec. (\ref{sig_nnnd}) and for the other
baryonic resonances we use eq. (\ref{fincrossexpr}) integrated over the 
resonance mass $\mu$ with $a$ and $b$ denoting the  
two nucleons in the initial state and $c$, $d$ and $e$  the $NN\pi$ 
final state. Introducing the proper isospin coefficients we obtain 
\bea
\sigma_{p\,p \to p\,p\,\pi^0}&=&\frac{2}{3} \sigma_{3/2}+\frac{1}{3} 
\sigma_{1/2}  \label{test1}
\\
\sigma_{p\,p \to p\,n\,\pi^+}&=&\frac{10}{3} \sigma_{3/2} +\frac{2}{3}
\sigma_{1/2}\\
\sigma_{p\,n \to p\, p\,\pi^-}&=&\frac{1}{3} \sigma_{3/2} +\frac{1}{3}
\sigma_{1/2}\\
\sigma_{p\,n \to p\, n\,\pi^0}&=&\frac{4}{3} \sigma_{3/2} +\frac{1}{3}
\sigma_{1/2}, 
\eea
with 
\bea
\label{sigma32}
\sigma_{3/2}&=&
\sum_{I_R=\frac{3}{2}}p\,p \to p\,R^+=
 \sigma_{p\,p \to p\,\Delta^+(1232)}\,
\nonumber
\\
&&+\frac{1}{4}
\frac{ 
\overline{\left| \cal M \right|^2}}
{16 \pi \,p_i \,s}
\, \int_{M_N+m_{\pi}}^{\sqrt{s}-M_N} \,d\mu \, p_f 
\times\sum_{R \neq \Delta(1232) \atop I_R=\frac{3}{2}}
\frac{\mu^2\, \Gamma_{R \to N\, \pi}(\mu)}
{ \left( \mu^2-M_R^2 \right)^2 +\mu^2\,\Gamma_{R,tot}^2(\mu)} \quad
\eea
and 
\bea 
\label{sigma12}
\sigma_{1/2}&=&\sum_{I_R=\frac{1}{2}}p \, p \to p\, R^+
\nonumber
\\
&=&
\frac{ 
\overline{\left| \cal M \right|^2}}
{16 \pi \,p_i \,s}
\, \int_{M_N+m_{\pi}}^{\sqrt{s}-M_N} \,d\mu \, p_f  
\sum_{R \atop I_R=\frac{1}{2}}
\frac{\mu^2\, \Gamma_{R \to N\, \pi}(\mu)}
{ \left( \mu^2-M_R^2 \right)^2 +\mu^2\,\Gamma_{R,tot}^2(\mu)}.
\eea
There is no explicit factor $\Gamma_{N\pi}$ in the contribution of the 
$\Delta(1232)$ to $\sigma_{3/2}$ (\ref{sigma32}) because  
it decays with $100 \%$ probability into a nucleon and a pion. \\
\item{$\rho$-production cross section}\\
The invariant matrix-elements for the production of baryon resonances which
can decay into a nucleon and a $\rho$ (cf. table \ref{tab_baryprop}) are
obtained by a fit to the experimental data for the reaction
\be
 pp \to pp \rho^0. \label{rhoprod}
\ee
Similar to eq. (\ref{test1}) we write the $\rho^0$-production cross section
as a sum of contributions of $I=3/2$- and $I=1/2$-resonances,
\be
\sigma_{pp \to pp\rho^0} = \frac{2}{3}\sigma_{3/2}
+ \frac{1}{3}\sigma_{1/2}, \label{deco1}
\ee
where
$\sigma_{3/2}$ and $\sigma_{1/2}$ are defined as in eqs. (\ref{sigma32})
and (\ref{sigma12}), while the sum extends only over resonances with
a non-vanishing branching ratio for the decay into a nucleon and a $\rho$.
The resulting matrix-elements are quoted in the second column of table
\ref{tab_baryprop}. The corresponding $\rho^0$-production cross section
is shown in fig. \ref{fig_resrho} in comparison to the experimental
data \cite{landolt}.
\begin{figure}
\vspace{-2cm}
\begin{center}
\hspace{-35mm}{\epsfig{file=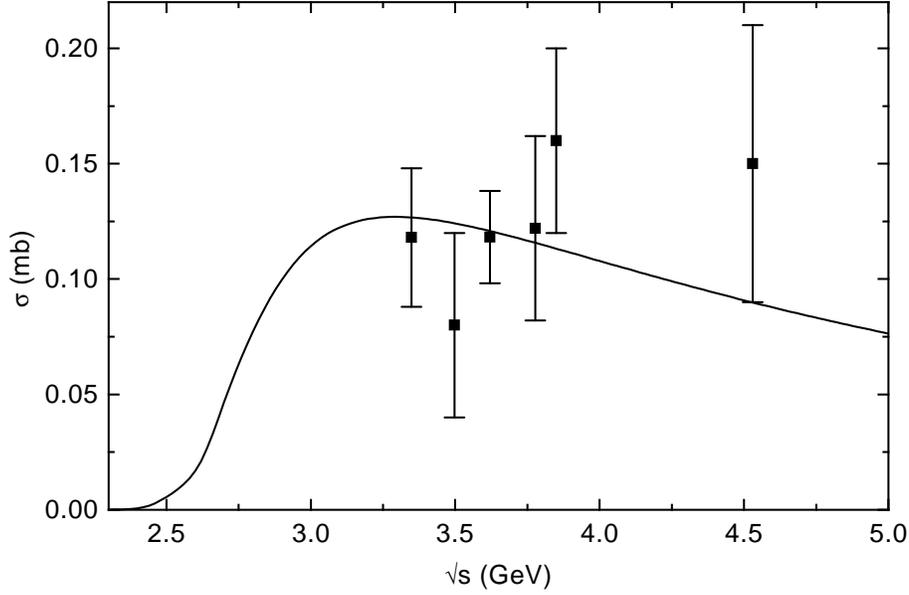,width=90mm}}
\vspace{-4.5cm}
\end{center}
\caption{Cross section for the reaction $pp \to pp\rho^0$ obtained from the
resonance model using the matrix elements given in table 
\protect\ref{tab_baryprop} (solid line) in comparison to the experimental data
\protect\cite{landolt}.}
\label{fig_resrho}
\end{figure}

\item{The $2\pi$-production cross section}\\
As already stated in sec. (\ref{sec_resprop}) the $2\pi$-production cross
section in nucleon-nucleon collisions is described via the excitation of
higher lying baryonic resonances and their subsequent decay into a nucleon
and two pions according to the branching ratios given in table 
\ref{tab_baryprop} and the corresponding decay widths:
\bea
1. \mbox{   } NN \to NR \to & N \Delta(1232) \pi & \to N N \pi \pi \label{rea1}\\
2. \mbox{   } NN \to NR \to & N N(1440) \pi       & \to N N \pi \pi \label{rea2} \\
3. \mbox{   } NN \to NR \to & N N \rho    & \to N N \pi \pi  \label{rea3}\\
4. \mbox{   } NN \to NR \to & N N \sigma          & \to N N \pi \pi. \label{rea4}
\eea
Here $R$ stands for the higher lying baryon resonances. In addition to these
processes we take into account the $2\pi$-production via the excitation of
two $\Delta(1232)$,
\bea
NN \to \Delta(1232) \Delta(1232) \to NN \pi \pi, \label{rea5}
\eea
adopting the  cross sections from ref. \cite{aichelin94}.
For the description of the $2\pi$-production cross section in nucleon-nucleon collisions
we define $\sigma^i_{3/2}$ and $\sigma^i_{1/2}$ ($i = 1,..,4$) similarly to eqs.
(\ref{sigma32}) and (\ref{sigma12}) replacing $\Gamma_{R \to N\pi}$ by
the corresponding decay width $\Gamma_i$ ($i = 1,..,4$) responsible for the
second step of the reactions (\ref{rea1}) to (\ref{rea4}).  The cross sections
then read
\bea
\sigma_{NN \to NN \pi \pi}= \sum_{i=1}^4 n_i \sigma^i_{1/2} +
\sum_{i=1}^4 d_i \sigma^i_{3/2} + \sigma_{NN \to \Delta(1232)
\Delta(1232) \to NN \pi \pi},
\label{cross2pifin}
\eea
where the $n_i$ and $d_i$ are the products of the isospin coefficients for the three 
steps of the reactions (\ref{rea1}) - (\ref{rea4}). 
The corresponding  factors are listed in table \ref{tab_2piclebsch}. 
\begin{table}[t]
\begin{center}
\begin{tabular}{|c||c|c|c|c|c|} 
\hline
\hline
& $pp \to pp\pi^+\pi^-$ &$pp \to pp\pi^0\pi^0$ & $pp \to pn\pi^+\pi^0$ & $pn \to pn\pi^+\pi^-$ & $pp \to pp\pi^-\pi^0$  \\
\hline
\hline  
$n_1$ & $5/9$   &       $2/9$       & $2/9 $ &  $5/9$ & $2/9$\\
\hline 
$n_2$ & $4/9$   &       $1/9$       & $4/9 $ &  $4/9$ & $4/9$\\
\hline 
$n_3$ & $1/3$   &       $  0$       & $2/3 $ &  $1/3$ & $2/3$\\
\hline 
$n_4$ & $2/3$   &       $1/3$       & $ 0   $ & $2/3$ & $       0   $\\
\hline 
$d_1$ & $26/45$ &       $2/45$      & $22/9 $ & $52/45$ & $17/45$\\
\hline 
$d_2$ & $2/9$   &       $2/9$       & $14/9 $ & $4/9$   & $5/9$\\
\hline 
$d_3$ & $2/3$   &       $ 0       $ & $10/3 $ & $2/3$   &       $1/3$\\
\hline 
$d_4$ & 0              & 0     & 0           &   0          & 0\\
\hline 
\hline
\end{tabular}
\end{center}
\caption{The products of the isospin coeficients for the three steps of the reactions 
(\protect\ref{rea1}) - (\protect\ref{rea4}) for those isospin channels where
 experimental data are available.}
\label{tab_2piclebsch}
\end{table}

\end{itemize}
Using the matrix elements already determined by the fits to the $\eta$- and 
$\rho$-production data we now adjust the unknown matrix elements 
$\overline{\left|\mbox{$\cal M$}_{NN\to NR} \right|^2}$ to reproduce the cross 
sections for $1\pi$- and $2\pi$-production in nucleon-nucleon collisions. 
\begin{figure}
\vspace{-2cm}
\begin{center}
\hspace{-45mm}{\epsfig{file=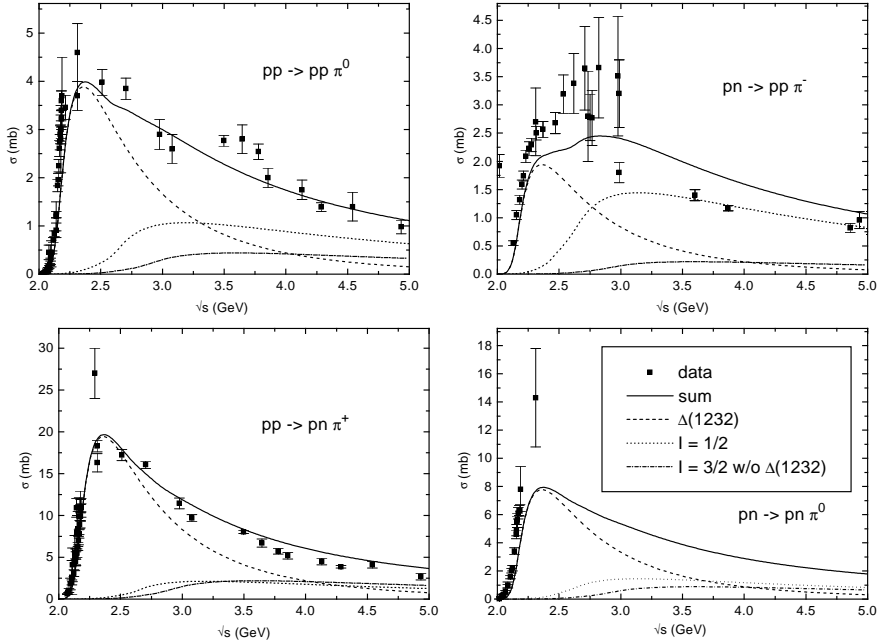,width=90mm}}
\vspace{-4.50cm}
\end{center}
\caption{The fitted $1\pi$-production cross sections (solid line) 
for different isospin channels in comparison to the 
data \protect\cite{landolt};
contributions from the $\Delta(1232)$ (dashed), 
the isospin 1/2 resonances (dotted), the higher isospin 3/2 resonances 
(dashed-dotted).}
\label{fig_vergl1pi}  
\end{figure}
The resulting $1\pi$-cross sections are shown in fig. \ref{fig_vergl1pi}
(solid line), where the contributions 
from the $\Delta(1232)$ (dashed line), the sum of all contributions 
from the isospin-$1/2$ resonances (dotted line) and the sum of all higher 
isospin-$3/2$ resonances (dash-dotted line) are displayed separately. 
Evidently the pion-cross 
sections are fitted in our resonance model 
reasonably well up to invariant energies of $5$ GeV.  \\
In fig. \ref{fig_twopi} we plot the resulting $2\pi$-production cross section 
(\ref{cross2pifin}) for the isospin channels where 
 experimental data  are
available \cite{landolt}. As in case of the $1\pi$-production channels 
the $2\pi$-data can be reproduced well within our multi-resonance model using the 
matrix elements given in table \ref{tab_baryprop}. 
\begin{figure}
\vspace{-2cm}
\begin{center}
\hspace{0mm}{\epsfig{file=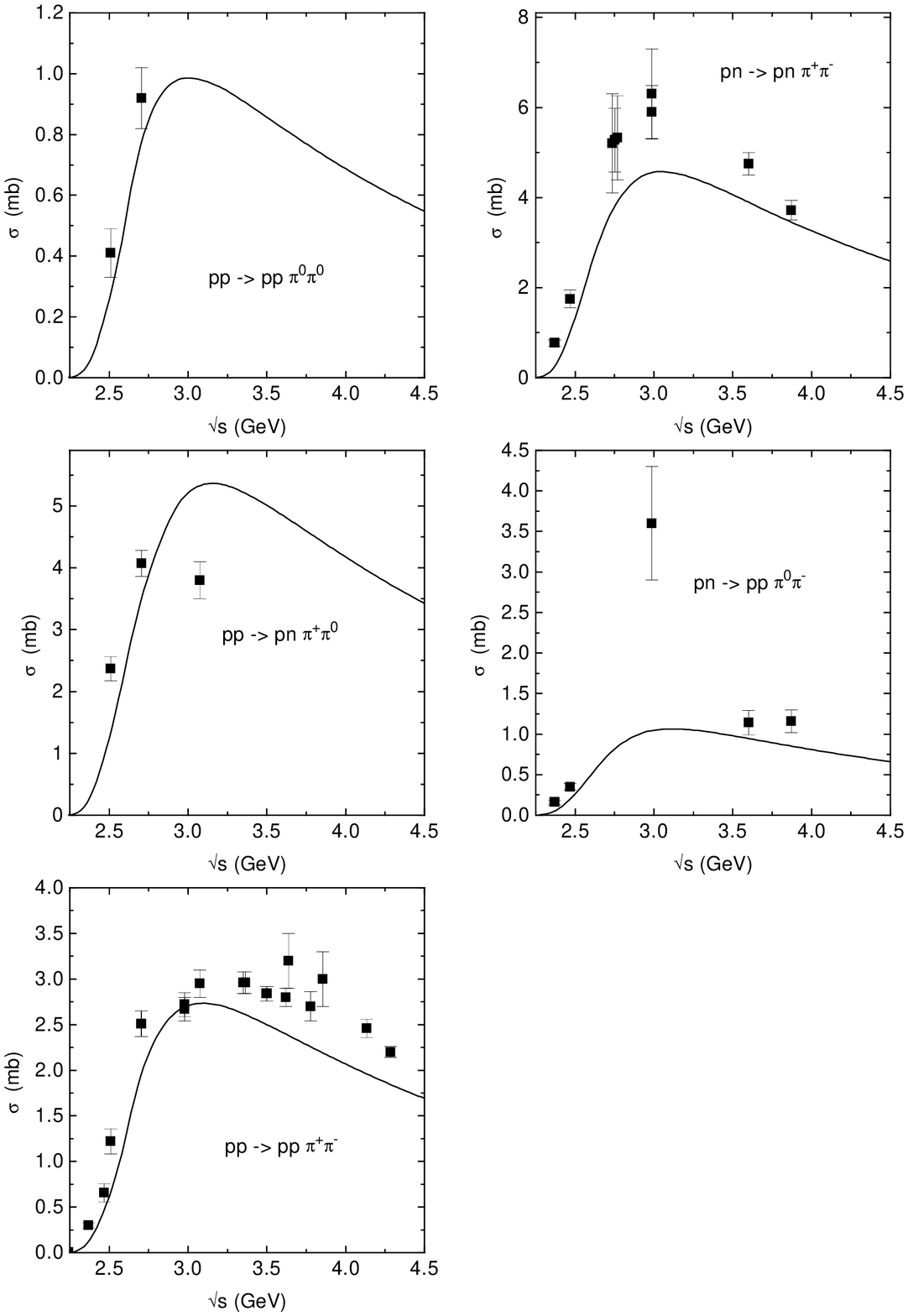,width=112mm}}
\vspace{-1cm}
\end{center}

\caption{Cross sections for $2\pi$-production in nucleon-nucleon collisions 
for different isospin channels obtained within the simple resonance model 
in comparison to the experimental data \protect\cite{landolt}.}
\label{fig_twopi}  
\end{figure} 
\end{subsubsection}
\begin{subsubsection}{The $NR\to NN$ Cross Section}\label{cros_nrnn}

The cross sections for the reaction $NR \to NN$, where 
$R$ stands for the $\Delta(1232)$ as well as a higher baryonic resonances, 
are given by an expression corresponding to eq. (\ref{crossgen}) for a two-body
final state. For the matrix elements involved we adopt 
those from the OBE-model of Dimitriev \cite{dimitriev86} for the
$\Delta(1232)$ and those                                 
from the simple resonance model (sec. (\ref{sig_nnnr})) in case of 
higher baryon resonances. Thus, in comparison to previous implementations 
of the BUU-model \cite{engel94,daniel91} we do not have to employ a 
detailed balance prescription to 
deduce the $NR \to NN$-cross section, since 
the matrix elements are known as a function of the invariant energy and the 
resonance mass. 
\end{subsubsection}
\begin{subsubsection}{The $NR \to NR'$-Cross Section}\label{sig_nrnrp}
For the collisions of a nucleon and a resonance $R$ leading to a nucleon and 
a different resonance $R'$ we use for all resonances (including the 
$\Delta(1232)$) the average of the matrix elements obtained for the reactions 
$NN \to NR$ and $NN \to NR'$. In analogy to eq. 
(\ref{fincrossexpr}) this leads to the following expression 
\bea 
\sigma_{NR \to NR^{\prime} \atop R \neq R^{\prime}} = & I\,
\frac{ 0.5(\left|\mbox{$\cal M$}_{NN \to NR}\right|^2 
+ \left|\mbox{$\cal M$}_{NN \to NR^\prime}\right|^2 )\,2\,(2J_{R^\prime}+1)  }{16\,\pi\,p_i\,s} \\ \nonumber  
& \times  \int \, d\mu \,p_f \frac{2}{\pi} \,
\frac{\mu^2 \, \Gamma_{R^{\prime}}(\mu)}{\left( \mu^2-M_{R^{\prime}}^2 \right)^2+
\mu^2\,\Gamma_{R^{\prime}}^2(\mu)},
\label{crossnrnrp}
\eea
where $I$ accounts for the proper isospin coefficients (cf. table
\ref{tab_iso2}) 
and $J_R'$ stands for the spin of the resonance in the final channel. 
\begin{table}[t]
\centerline{
\begin{tabular}{|ll|ll|c|} 
\hline
 & & & &I\\
\hline
$N^+$ & $N^+$ & $N^+$ & $N^+$ & 1\\
\hline
$N^+$ & $N^0$ & $N^+$ & $N^0$ & 1/2\\
\hline
$N^+$ & $N^+$ & $N^0$ & $\Delta^{++}$ & 3/4\\
\hline
$N^+$ & $N^+$ & $N^+$ & $\Delta^+$ & 1/4\\
\hline
$N^+$ & $N^0$ & $N^+$ & $\Delta^0$ & 1/4\\
\hline
$N^+$ & $\Delta^{++}$ & $N^+$ & $\Delta^{++}$ & 1\\
\hline
$N^+$ & $\Delta^+$ & $N^0$ & $\Delta^{++}$ & 3/8\\
\hline
$N^+$ & $\Delta^+$ & $N^+$ & $\Delta^+$ & 5/8\\
\hline
$N^+$ & $\Delta^0$ & $N^+$ & $\Delta^0$ & 1/2\\
\hline
$N^+$ & $\Delta^-$ & $N^+$ & $\Delta^-$ & 5/8\\
\hline
\end{tabular}}
\caption{Isospin coefficients $I$ for baryon-baryon collisions. $N$ and $\Delta$ represent 
isospin $1/2$ and isospin $3/2$ particles, respectively. 
The isospin channels not listed
explicitely are determined by assuming isospin symmetry.}
\label{tab_iso2}
\end{table}

\end{subsubsection}
\begin{subsubsection}{The $NN\to NN\pi$-Cross Section}\label{sig_nnpi}
The $NN \to NR$-cross sections are determined by fitting the experimental 
data for $1\pi$-production in nucleon-nucleon collisions above the 
$\Delta(1232)$. As compared to the 
data the resulting $1\pi$-cross sections  are slightly too low just above the 
$\pi$-production threshold (cf. fig. \ref{fig_thresh}). 
\begin{figure}
\vspace{-2cm}
\begin{center}
\hspace{-45mm}{\epsfig{file=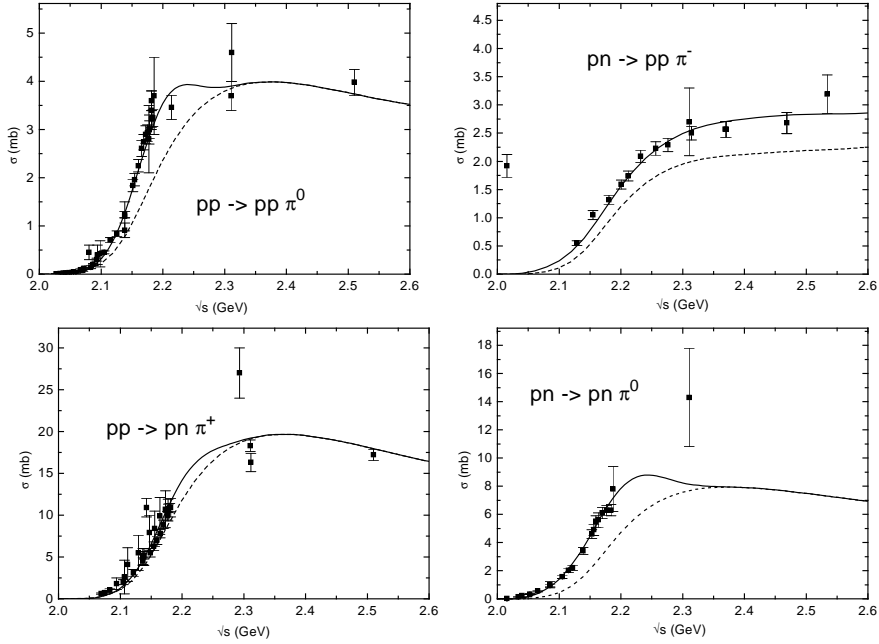,width=90mm}}
\vspace{-4.5cm}
\end{center}
\caption{The $1\pi$-production cross section taking into account only the 
contributions due to the decay of baryonic resonances (dashed line) and 
the total cross sections obtained by adding the direct pion 
production in nucleon-nucleon collisions (solid line) in  comparison to the 
experimental data \protect\cite{landolt}.}
\label{fig_thresh} 
\end{figure}
In order to compensate for this deficit we attribute the difference 
between the data and the cross sections obtained from the resonance  
model to the cross section for direct (s-wave) $\pi$-production in nucleon-nucleon 
collisions ($NN\to NN\pi$). 
The resulting difference (s-wave) cross section can be fitted by 
the expression 
\bea
\label{fitunter}
\sigma_{N\,N\to N\, N\,\pi}(x)=A \, x^{n_1} \, 
e^{-\left( a\,x^{n_2} + b\,x \right)}, 
\eea
with 
\bea
 x=\frac{\sqrt{s}-2M_N-m_{\pi}}{5\,{\rm GeV}}, \non
\eea 
using the parameters given in table \ref{tab_fit}. 
\begin{table}
\centerline{
\begin{tabular}{|l||r|r|r|r|r|} 
\hline
channel& A [mb] & a & b & $n_1$& $n_2$\\
\hline
\hline
$p \quad p\to p\quad p \quad \pi^0$& 61.3 & 1.52& 2.50 & 6.18 & 3.48\\
\hline
$p \quad p\to p\quad n \quad \pi^+$& 122.6 & 1.52 & 2.50 & 6.18 & 3.48\\
\hline
$p \quad n\to p\quad p \quad \pi^-$& 24.9 & 3.30 & 0.85 & 1.93 & 0.002\\
\hline
$p \quad n\to p\quad n \quad \pi^0$& 7.25 & 0.88 & 0 & 2.31 & 3.64\\
\hline
\end{tabular}}
\protect\caption{Parameters obtained for the direct $\pi$-production 
cross section $\sigma_{N\,N \to N\,N\,\pi}$ (eq. (\protect\ref{fitunter})).}
\label{tab_fit}  
\end{table}
Now adding $\sigma_{N\,N \to N\,N\,\pi}$ incoherently to the $1\pi$-production 
cross section from the baryon resonance decays then yields the total cross 
section depicted in fig. \ref{fig_thresh} by the solid line which now also 
gives a fit at threshold. 
\end{subsubsection}
\begin{subsubsection}{The $NN\pi \to NN$-Rate}\label{sig_pitrans}
Assuming that the invariant matrix element for the reaction 
$N_1 + N_2 \to N_3 + N_4 + \pi$ depends only on the invariant energy one 
can write the cross section $\sigma_{N_1\,N_2 \to N_3\,N_4\,\pi}$ (see sec. 
(\ref{sig_nnpi})) in the following form \cite{pdg} 
\bea
\sigma_{N_1N_2 \to N_3N_4 \pi}(\sqrt{s}) = 
\frac{S_{N_1,N_2}}{64 \,(2\,\pi)^3 \,p_i\, \sqrt{s}^3 } \,  
\left|\mbox{$\cal M$}_{N_1\,N_2 \to N_3\,N_4\,\pi}(\sqrt{s})\right|^2 
\,\int dm_{34}^2 \, dm_{3\pi}^2.
\label{transpi}
\eea 
Here $S_{N_1,N_2}$ is the symmetry factor for $N_1$ and $N_2$, 
$p_i$ is the CMS momentum of the nucleons in the initial state and 
\bea 
m_{34}^2 = (p_3 + p_4)^2, \mbox{      } m_{3\pi}^2 = (p_3 + p_\pi)^2. 
\non
\eea
The transition rate for a pion being absorbed by $N_3$ and $N_4$ is given by
\bea
W_{fi}=(2\pi)^4 \, \frac{\delta^4 \left( p_{N,1}+p_{N,2}-p_{N,3}-p_{N,4}-
p_{\pi} \right)\,|\mbox{$\cal M$}|^2}{V^5} \quad, \label{transallg}
\eea
where the normalization volume $V$ contains $2E$ particles \cite{halzen}. 
Multiplying this equation with the phase-space factors for the two nucleons 
($N_1$, $N_2$) in the final state and taking into account that the pion 
reacts with two nucleons from the surrounding nuclear medium gives 
\bea
\Gamma_{\pi N_3 N_4 \to N_1 N_2}&=& 
 S_{N_1,N_2}\,\frac{p_f}{4\,\pi\,\sqrt{s}} |\mbox{$\cal M$}|^2
\,\frac{1}{2\,E_{\pi}}\,\frac{\rho_{N,3}}{2E_{N,3}}\,
\frac{\rho_{N,4}}{2E_{N,4}},
\eea
where $\rho_{N,3}$ and $\rho_{N,4}$ are the corresponding local neutron or 
proton densities.
\end{subsubsection}
\end{subsection}
\begin{subsection}{Elastic Baryon-Baryon Cross Sections}\label{sig_elast}
For the elastic nucleon-nucleon cross section we use the 
conventional Cugnon parametrization \cite{bertsch88,cugnon82} 
\begin{equation}
\sigma_{NN \to NN}=\left(\frac{35}{1+\frac{\sqrt{s}-2M_N}{{\rm GeV}}}+
20\right)\,{\rm mb}.
\label{cugnonpara}
\end{equation}
The cross sections for nucleon-resonance scattering (with the same 
baryonic resonance in the initial and final channel) are determined  
in the following way: assuming an isotropic 
angular dependence for (\ref{cugnonpara}) we fix the matrix element 
squared by 
\begin{equation}
\left| \mbox{$\cal M$}_{NN \to NN} \right|^2=16\,\pi\,s\left(
\frac{35}{1+\frac{\sqrt{s}-2M_N}{{\rm GeV}}}+20\right)\,{\rm mb}. 
\label{elastmatri}
\end{equation}
Now inserting (\ref{elastmatri}) in eq. (\ref{fincrossexpr}) we obtain for  
the elastic nucleon-baryon scattering cross section 
\bea
\sigma_{NR \to NR} = \frac{\left|\mbox{$\cal M$}_{NN \to NN}\right|^2}{16\,\pi\,p_i\,s} 
\int \, d\mu \,p_f \frac{2}{\pi} \,
\frac{\mu^2 \, \Gamma_R(\mu)}{\left( \mu^2-M_R^2 \right)^2+\mu^2\,\Gamma_R^2(\mu)}.
\label{crossnrnr}
\eea
We note that
the cross section (\ref{crossnrnr}) in addition to elastic scattering 
also allows for a change in mass of the resonance in the scattering process. 
\end{subsection}
\end{section}
\begin{section}{Results for Nucleus-Nucleus Collisions 
and Comparison to Data}\label{sec_res}
As described above the current 
implementation of the CBUU model takes into account baryonic resonances 
up to masses 
of $1.95$ $GeV/c^2$ for the first time. Hence, we start our discussion 
with an analysis of the effects of these resonances on the resulting 
pion spectra. Since the higher baryon resonances  
decay ultimately into a nucleon and one pion or into a nucleon and 
two pions (cf. sec. (\ref{sig_nnnr})) we expect the contributions from those
resonances to show up in the 
low energy as well as in the high energy regime of the pion spectra. 
An enhancement of the pion yield in the low energy regime is expected due 
to the $2\pi$-channels ($NN \to \Delta(1232) \Delta(1232) \to NN \pi \pi, \,
NN \to NR \to N \pi \pi$, sec. (\ref{sig_nnnr})). Pions produced  
via the decay of two $\Delta(1232)$, that are excited simultaneously in a
nucleon-nucleon collisions, or via the $2\pi$-decay of the higher resonances 
have a lower momentum than those stemming from the reaction $NN \to 
NR \to NN \pi$. Thus we expect an enhancement of the low energy 
pion yield. On the other hand, once a higher baryon resonance is 
excited and subsequently decays into a nucleon and a single pion, this pion will 
have a higher momentum in the restframe of the resonance than those 
emitted in $\Delta(1232)$-decays. 

In fig. \ref{fig_twopi1} we show the effect of the $2\pi$-channels on the 
low energy regime of the pion spectra. Fig. \ref{fig_twopi1} displays the 
total $\pi$-multiplicity weighted with $1/p_T$ as a function of the pion transverse 
momentum $p_T$ for a central $Au + Au$ reaction at $1.5$ AGeV. The solid 
line represents the result of a full CBUU-calculation and the dashed line
is obtained by switching off the $2\pi$-channels. This is done by neglecting 
the cross section $NN \to \Delta(1232) \Delta(1232)$ as well as that for 
the time reversed reaction. The effect of the $2\pi$-decay of the higher resonances 
is eliminated by allowing only the decay into a nucleon and a single pion,
i. e. 
($\Gamma_{1\pi}/\Gamma_{tot} = 1$ and $\Gamma_{2\pi} = 0$). In addition the 
matrix elements for resonance production in 
nucleon-nucleon scattering have been refitted in order to guarantee the correct 
description of the $1\pi$-production cross sections in nucleon-nucleon 
collisions. The resulting value for the matrix element squared in the latter
case is 
\bea
\overline{\left|\mbox{$\cal M$}_{N,N\to N,R} \right|^2} = 
16 \pi \times 9 \,{\rm mb \, GeV^2}. \nonumber 
\eea
\begin{figure}[htb]
\vspace{-2cm}
\begin{center}
\hspace{-18mm}{\epsfig{file=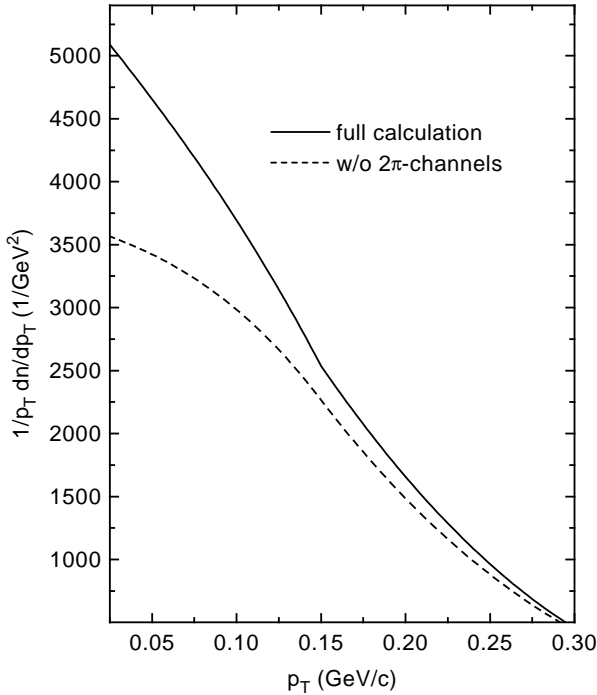,width=95mm}}
\vspace{-2.0cm}
\end{center}
\caption{Total pion multiplicity weighted with $1/p_T$ as a function of the
transverse momentum $p_T$ for  
a central ($b = 0.0$ fm) $Au + Au$ collision at $1.5$ GeV/A. The solid 
line shows the result of the full CBUU-calculation and the dashed line 
corresponds to the result when omitting the $2\pi$-production channels 
as described in the text.}  
\label{fig_twopi1}
\end{figure}
As can be seen from fig. \ref{fig_twopi1} at 1.5 GeV/A the $2\pi$-channels 
increase the pion yield at low $p_T$ by $20 - 25 \%$. The effect vanishes 
already at a transverse momentum of $0.25$ to $0.3$ GeV/c. For beam energies 
about 1.0 GeV/A we find that this effect is reduced to $\approx 10 \%$. \
\begin{figure}[htb]
\begin{center}
\hspace{-45mm}{\epsfig{file=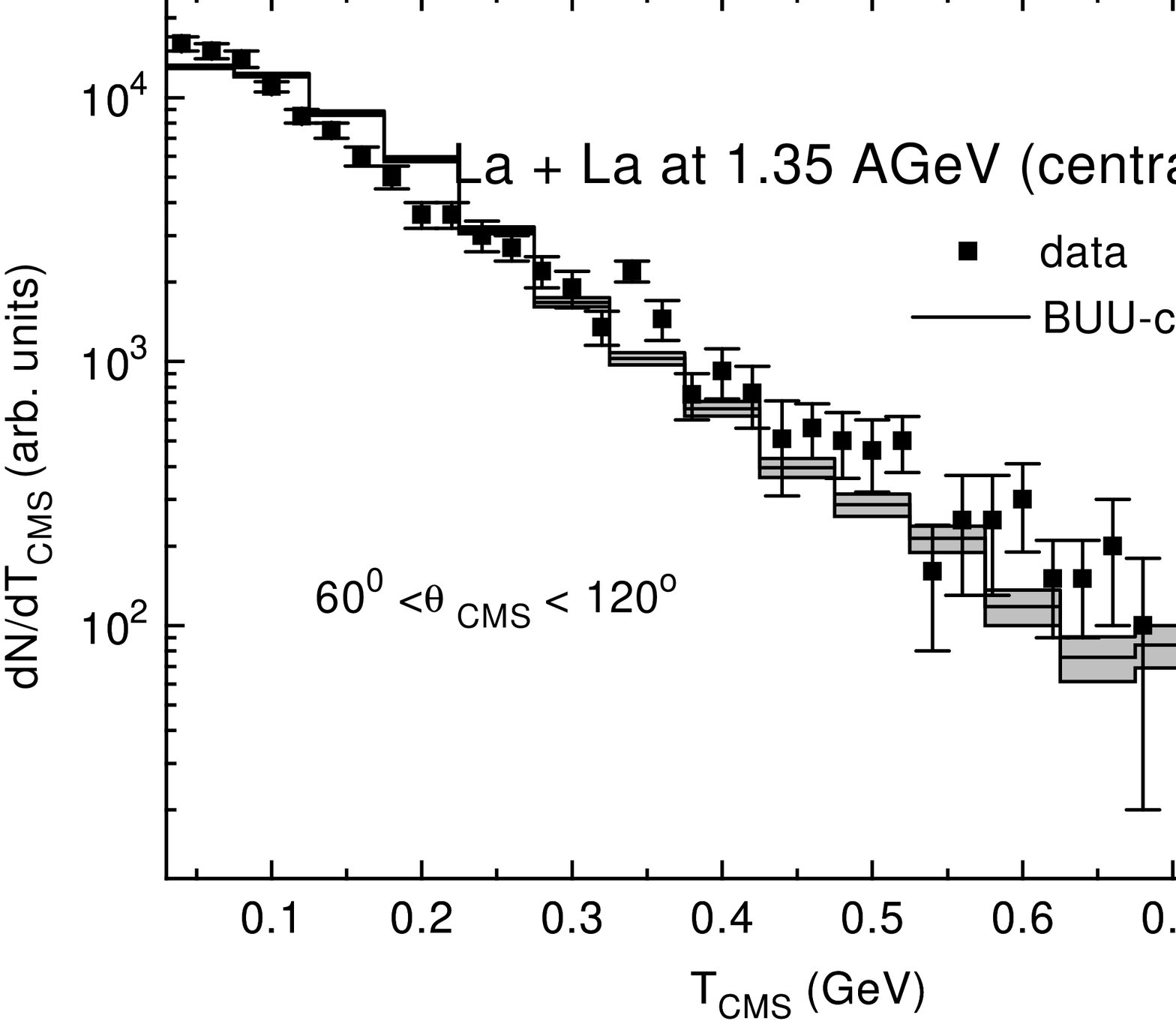,width=95mm}}
\vspace{-4.5cm}
\end{center}
\caption{The $\pi^-$-multiplicity for central $La + La$ collisions at  
$1.35$ GeV/A as a function of the center of mass pion kinetic energy for 
$80^o \le \Theta_{CMS} \le 120^o$. The solid histogram displays the 
result of the calculation while the shaded areas indicate the statistical 
error of the calculation. The data (squares) are taken from 
\protect\cite{odyn87}.}  
\label{fig_lala1}
\end{figure}
%
%
\begin{figure}[htb]
\vspace{-2cm}
\begin{center}
\hspace{-45mm}{\epsfig{file=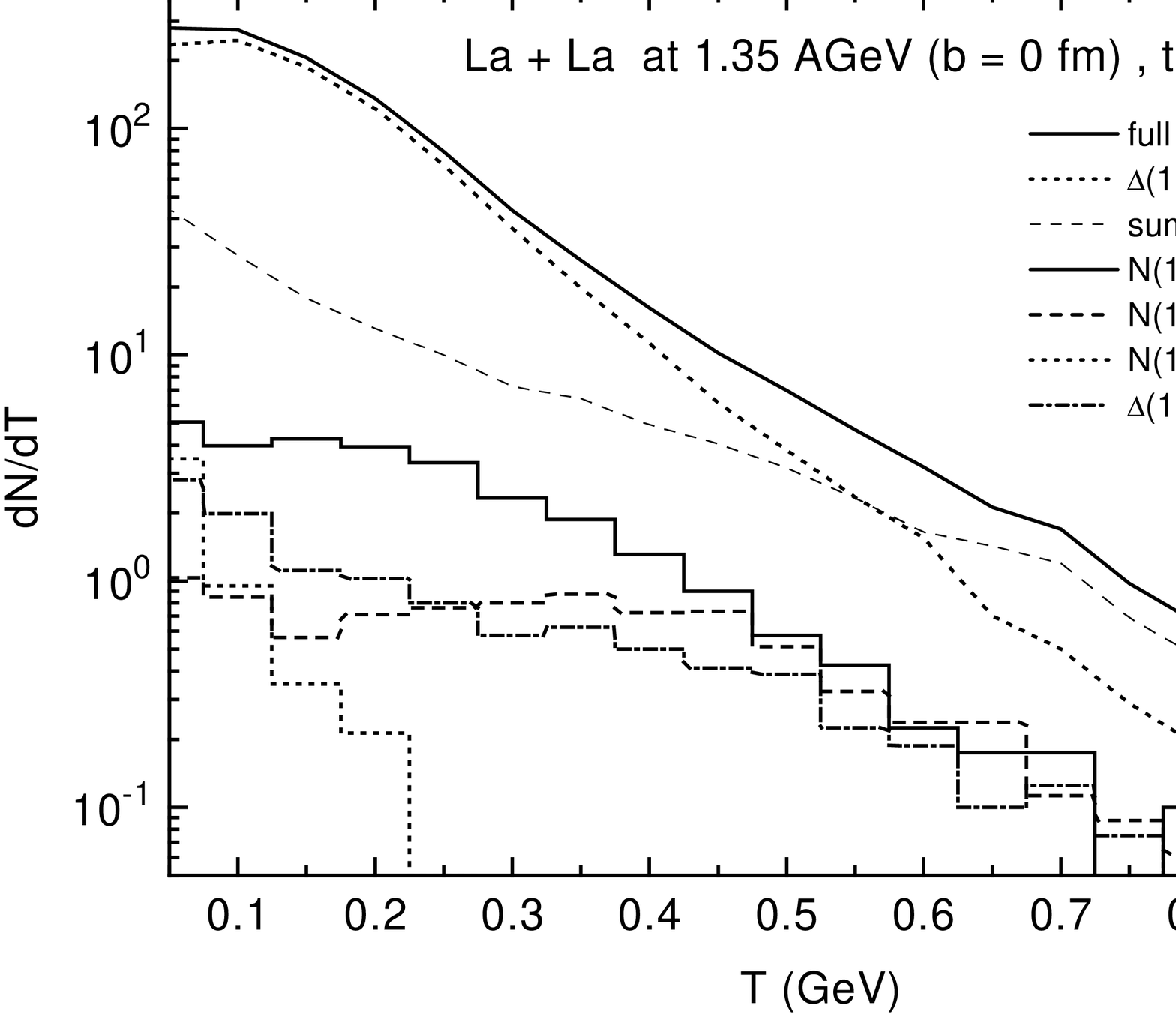,width=95mm}}
\vspace{-4.5cm}
\end{center}
\caption{The $\pi^-$-multiplicity for central $La + La$ collisions at  
$1.35$ AGeV as a function of the center-of-mass pion kinetic energy.  
Solid line: CBUU-calculation, dotted line: fraction of pions stemming 
from $\Delta(1232)$-, $\rho$-, $\sigma$-decay or from direct pion 
production in nucleon-nucleon collisions; dashed line: fraction of 
pions originating from decays of higher baryonic resonances. Histograms: 
fraction of pions from specific resonance decays: solid $N(1440)$, 
dotted $N(1535)$, dash-dotted $N(1520)$ and dashed $\Delta(1600)$.}  
\label{fig_lala3}
\end{figure}
After investigating the effect of the $2\pi$-channels on the $\pi$-spectra 
we now discuss the impact of the higher baryonic resonances at high 
pion kinetic energies. In this respect we display in fig. \ref{fig_lala1} the 
$\pi^-$-multiplicity in central $La + La$ collisions at 1.35 AGeV as a 
function of the center-of-mass kinetic energy of the pions at 
$80^o \le \Theta_{CMS} \le 120^o$ obtained from a CBUU-calculation 
in comparison to the experimental data from \cite{odyn87}. 
We are able to reproduce the experimental spectrum; 
especially the slope at high pion energies 
($T_{CMS} \ge 0.5$ GeV) is described well by the calculation. 
That this slope is influenced significantly by the higher lying resonances 
is shown in fig. \ref{fig_lala3}, where we display the calculated 
pion multiplicity for a central $La + La$ collision at 1.35 GeV/A as a 
function of the center-of-mass pion kinetic energy. The upper solid line 
corresponds to the full calculation while the dotted line displays final pions 
 stemming from direct pion production in nucleon-nucleon collisions or 
from $\Delta(1232)$-, $\rho$- or $\sigma$-decay. 
The dashed line corresponds to the sum 
of the final pions stemming from the decay of higher baryonic resonances; 
the histograms show individually the contributions from single resonances to 
the pion yield: $N(1440$ (solid line), $N(1535)$ dashed line, $N(1520)$ 
dash-dotted line and $\Delta(1600)$ dashed line. The sum of the contributions 
from higher resonances for low kinetic energies is in the order of $10 - 20\%$ 
while the yield above $T_{CMS} \approx 0.6$ 
GeV is fully dominated by pions originating 
from the decay of higher resonances. \\

We now turn to a comparison of our calculations to the data on $\pi^-$-production 
in $Ar$ + $KCL$ collisions at $1.8$ GeV/A obtained at the BEVALAC. All data
shown in the following are taken from ref. \cite{stock86}. In the experimental 
analysis a central and a minimum bias event class have been used. In order 
to be able to relate the CBUU-results to the experimental events we determine these 
event classes by comparing to the $\frac{d\sigma}{dp_T}$-spectrum. 
The inclusive cross section is obtained by the integration of the 
multiplicities $n(b)$ over the impact parameter $b$ via 
\be
\frac{d\sigma}{dp_T} = 2\pi \, \int_0^\infty \frac{dn(b)}{dp_T} b db.
\label{sec_def}
\ee
In order to identify the "central event" class we thus integrate up to 
a maximal impact parameter $b_{max}$, which we 
determine by fitting the data. Fig. \ref{fig_datapt} shows these cross 
sections as a function of the transverse pion momentum $p_T$ from  
the CBUU-model for $b_{max}$ from $1.6$ to $2.3$ $fm$. In this 
$b_{max}$-regime one can see that the shape of the 
calculated spectrum does not 
depend on the events used; one only observes a shift of the spectrum in magnitude 
when including more events. Since the integrated spectrum is reproduced well 
for $b_{max} = 2.1$ $fm$, we will use events with $b \leq 2.1 fm$
for the comparison of the 
CBUU-calculations with the data.

\begin{figure}
\vspace{-3cm}
\begin{center}
\hspace{-35mm}{\epsfig{file=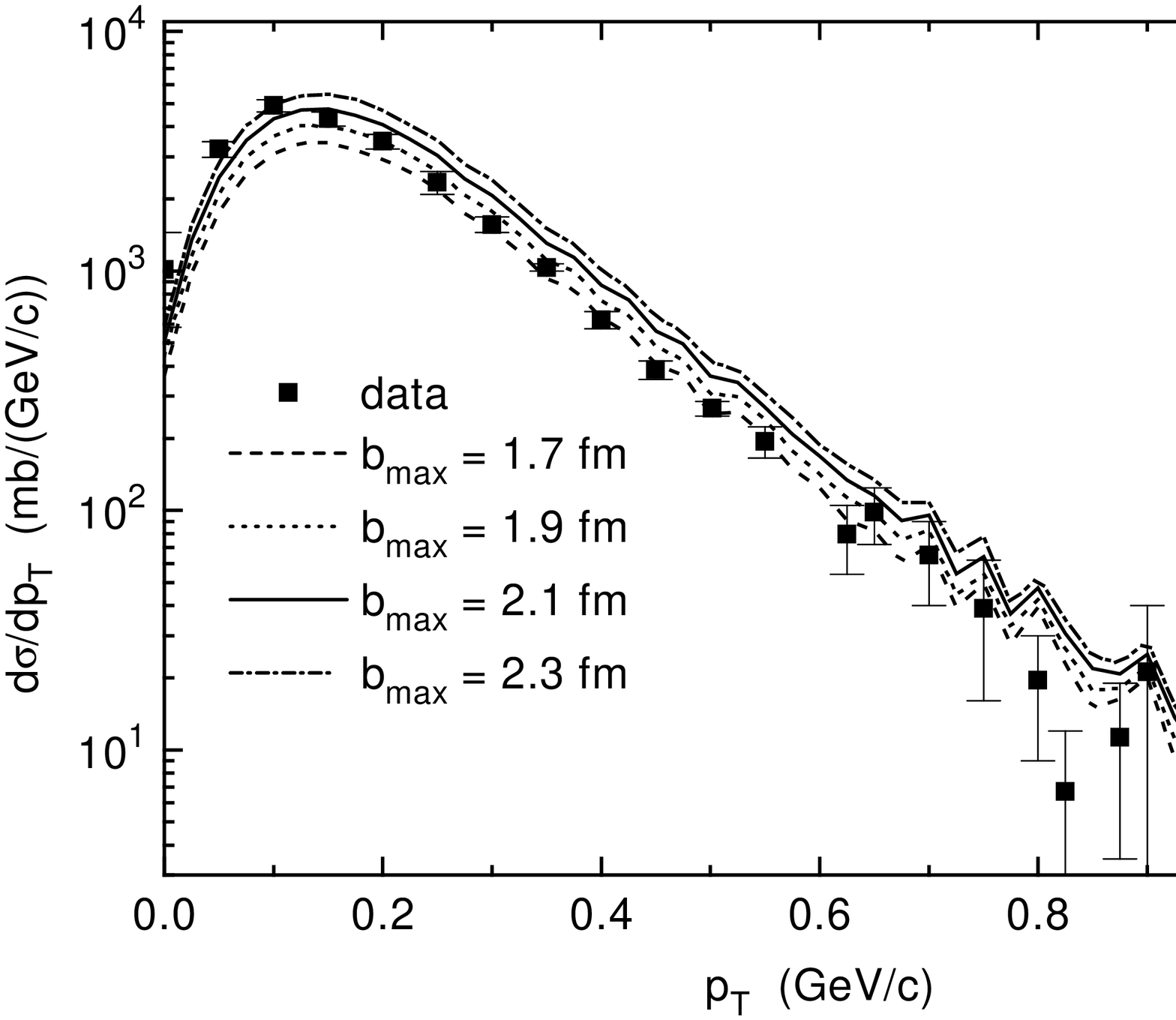,width=95mm}}
\vspace{-4.5cm}
\end{center}
\caption{The $\pi^-$-cross section for central event classes as 
a function of the transverse momentum $p_T$. The squares indicate 
the data from \protect\cite{stock86} while the lines show the result of 
a CBUU-calculation with different $b_{max}$ 
: solid line $b_{max}= 2.1$ fm , dashed line $b_{max}= 1.7$ fm, 
dotted line $b_{max}= 1.9$ fm, dash-dotted line $b_{max}= 2.3$ fm.}  
\label{fig_datapt}
\end{figure}
\begin{figure}
\vspace{-1cm}
\begin{center}
\hspace{-35mm}{\epsfig{file=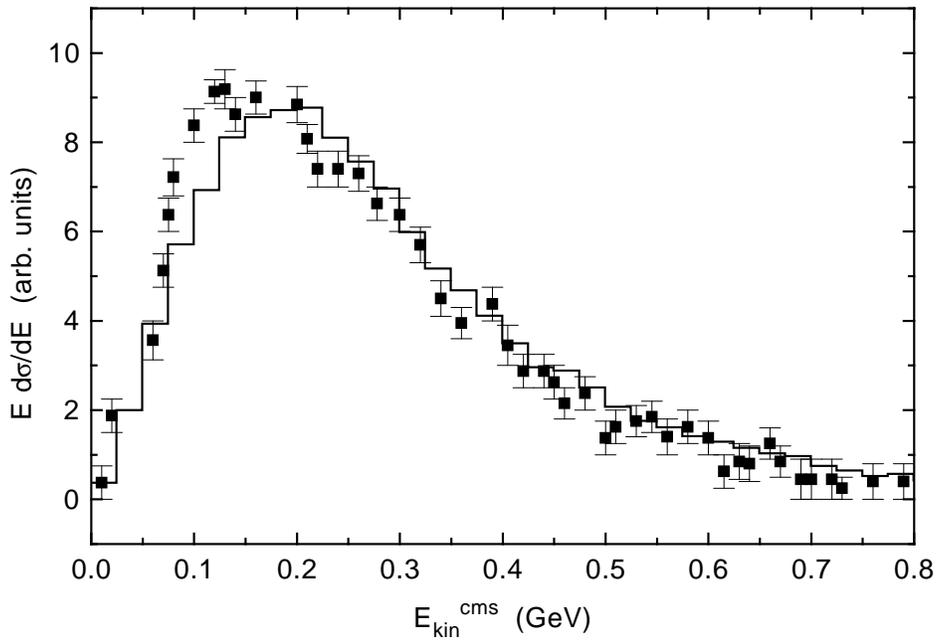,width=95mm}}
\vspace{-4.5cm}
\end{center}
\caption{The differential cross section $E\frac{d\sigma}{dE}$ 
for the central event class ($b_{max}=2.1 fm$) as a function of the pion-kinetic energy. The squares indicate the data 
from \protect\cite{stock86} and the solid line corresponds to the 
calculation.}
\label{fig_dataedsde}
\end{figure}
\begin{figure}
\vspace{-1cm}
\begin{center}
\hspace{-35mm}{\epsfig{file=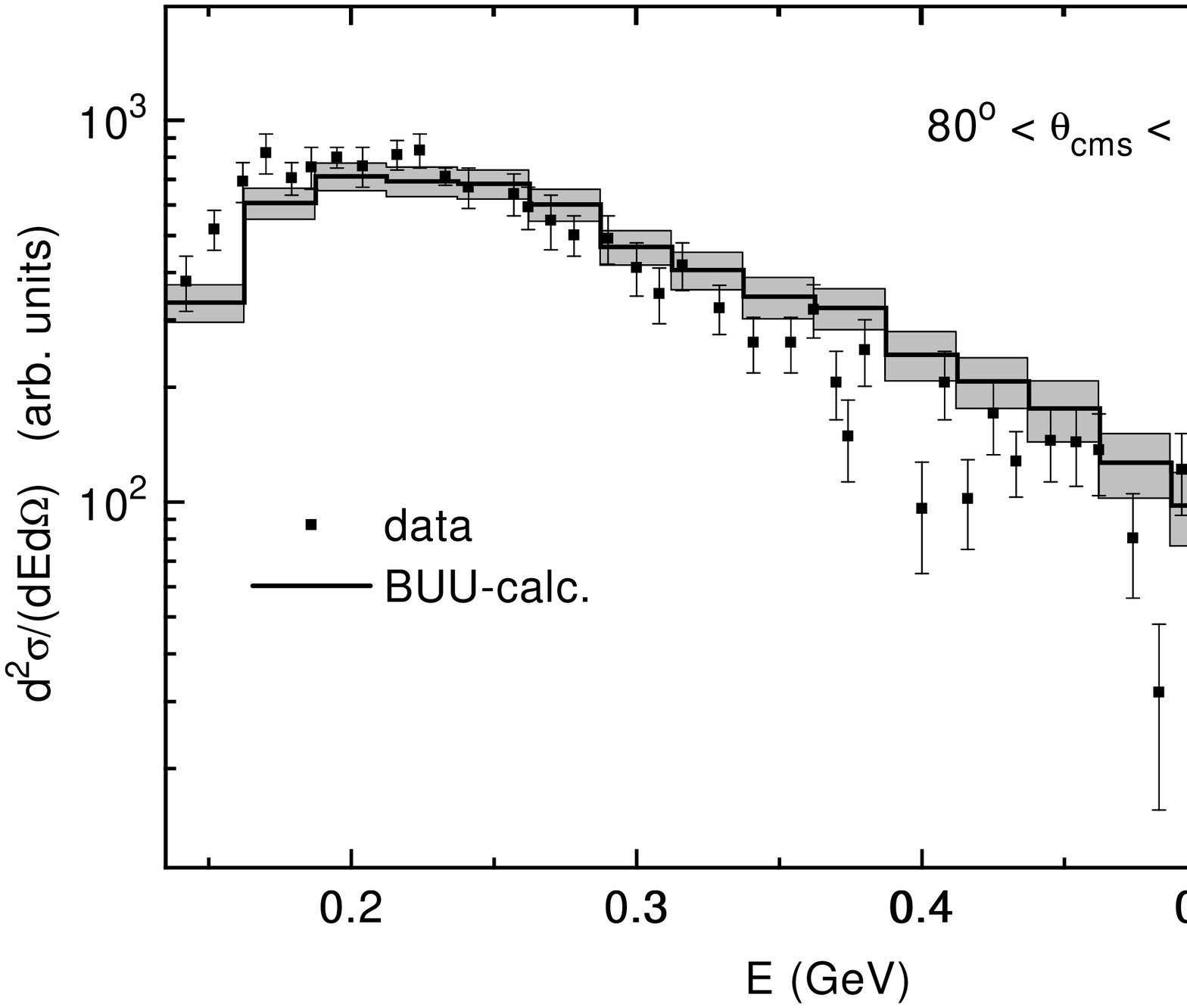,width=90mm}}
\vspace{-4.5cm}
\end{center}
\caption{The double differential $\pi^-$-cross section for the central 
event class as a function of the pion-energy. The squares represent the 
data, while the solid line corresponds to the result of the CBUU-calculation. 
The shaded areas indicate the statistical errors of the calculation.}
\label{fig_dataedsdedo}
\end{figure}
From fig. \ref{fig_datapt} 
we see that the spectral form of the data is described well over the 
whole momentum range except for the transverse momenta between $0.3$ and 
$0.55$ GeV/c where the data are overestimated. This overall agreement 
is also found when looking at 
$E\frac{d\sigma}{dE}$ as a function of the pion kinetic energy in the 
CMS (cf. fig. \ref{fig_dataedsde}), although we find that the distribution  
resulting from the calculation seems to be shifted by $\approx 25$ MeV to higher 
energies. In fig. \ref{fig_dataedsdedo},  
furthermore, we show a double differential spectrum 
where the data are indicated by the squares and the 
solid histrogram corresponds to the result of the CBUU-calculation  
obtained by scaling the original calculation in units  
$mb/GeV/sr$ by a factor $1.9$. Again we observe a good agreement for the 
form of the spectrum, although again the calculation seems to overestimate 
the data between 0.3 and 0.5 GeV. \\
\\
After convincing ourselves that the CBUU-model reproduces the spectral form 
for the $\pi^-$-distributions rather well we now turn to differential pion 
angular distributions. In fig. \ref{fig_datadndcos} we display 
$\frac{dN}{dcos\theta}$ for the inclusive 
$\pi^-$-yield, i.e. integrating eq. (\ref{sec_def}) 
up to $\infty$ (top), and for the central event class (bottom) in comparison 
to the data \cite{stock86}. The circles represent the data while the 
squares indicate the results of the calculation. The solid lines are 
fits to the calculations employing the functional form \cite{stock86}
\be
\frac{dN}{dcos\theta} = const \times (1.0 + a\, cos^2 \theta).
\label{def_aniso}
\ee
\begin{figure}
\begin{center}
\vspace{-1cm}
\epsfig{file=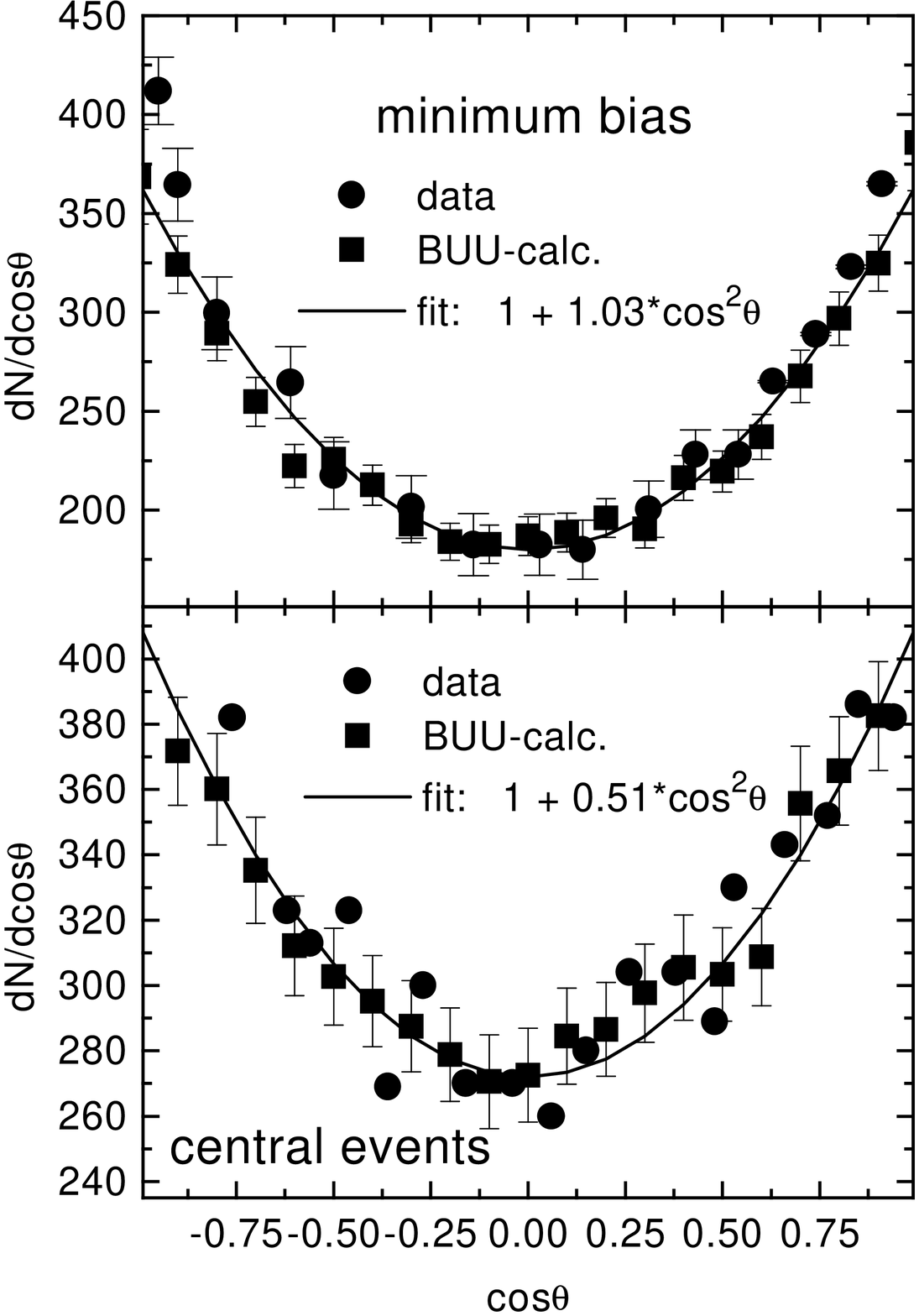,width=100mm}
\vspace{-1.5cm}
\end{center}
\caption{The number of pions as a function of $cos\theta$ in the 
CMS for minimum bias (top) and central (bottom) events. The data are 
represented by the circles and the results of the CBUU-calculations 
are given by the squares. The solid lines represent fits of the form 
(\protect\ref{def_aniso}) to the CBUU-calculations.}
\label{fig_datadndcos}
\end{figure}
For both event classes the data are
reproduced remarkably well. When considering the minimum bias event class 
we find for the anisotropy parameter $a \approx 1.03$, while $a$  
is reduced to $\approx 0.51$ when looking only at central events. As stated 
in \cite{stock86} this decrease of anisotropy - when going from minimum bias 
to central events - can be understood as an effect of the centrality of 
the heavy-ion collision because for minimum bias events semi-peripheral 
and peripheral collisions are weighted stronger than central collisions (c.f. 
eq. (\ref{sec_def})). Pions produced in semi-peripheral and peripheral 
collsions are more likely to originate from first chance 
$NN \to N\Delta \to NN \pi$ collisions than those produced in central 
collisions. Thus the anisotropy introduced in these first chance collisions 
prevails over the more isotropic pion distributions originating from central 
collisions. This line of argument is supported by the fact that the anisotropy 
coefficient is reduced to half of its value when considering only 
central collisions (see. bottom of fig. \ref{fig_datadndcos}). In central  
heavy-ion collisions a high density regime is formed where pions and baryonic 
resonances are produced and absorbed repeatedly. Thus, pions from central
collisions result mainly from multi-step processes and hence are emitted more 
isotropically than pions from peripheral collisions. \\
Finally, we look (for the 
central event class) at the energy dependence of the anisotropy parameter $a$ assuming 
as in \cite{stock86}
\be
\sigma(E) \sim  1 + a(E)\, cos^2 \theta. \label{def_aniso_e}
\ee
\begin{figure}[tb]
\vspace{-2cm}
\begin{center}
\hspace{-35mm}{\epsfig{file=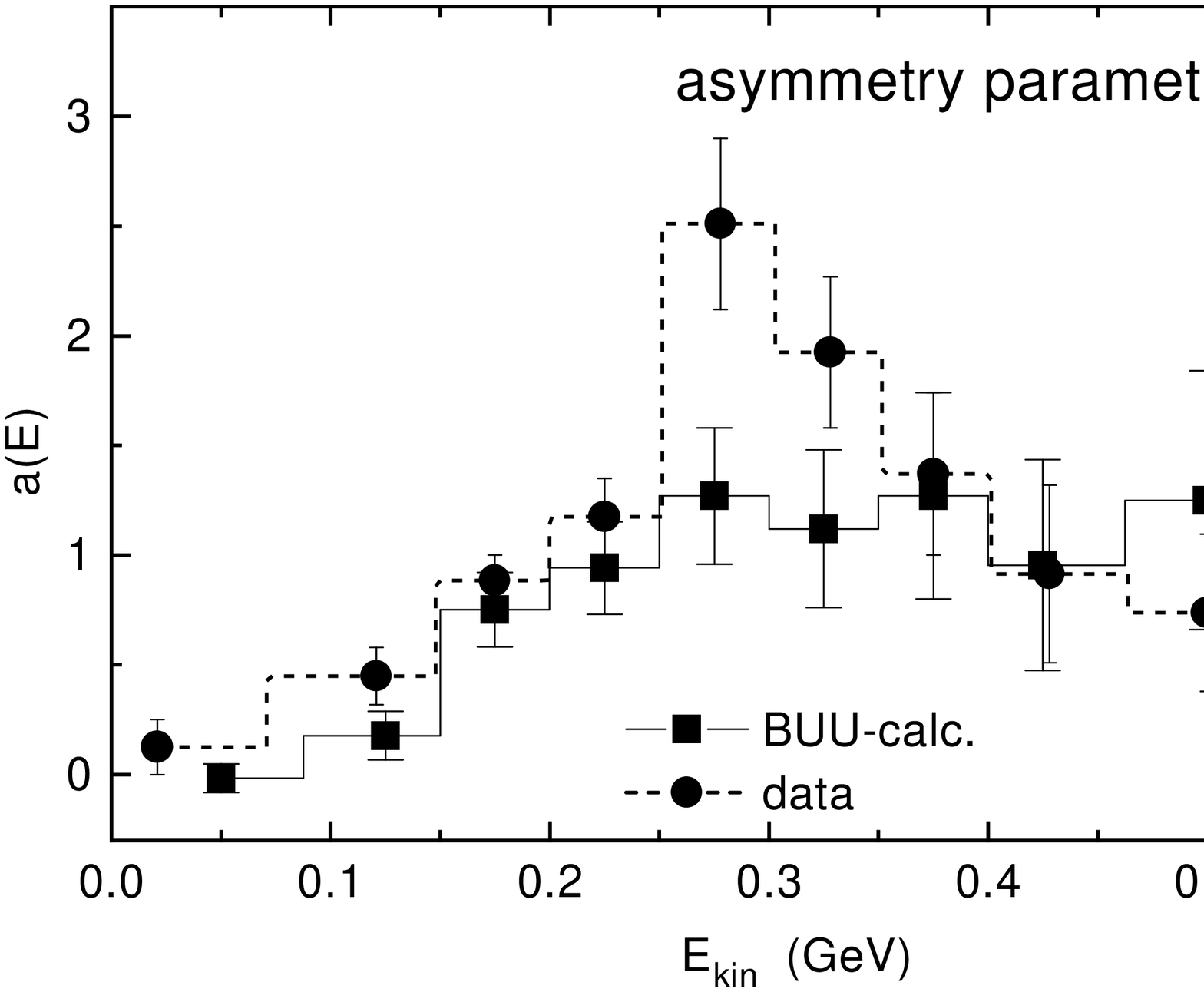,width=90mm}}
\end{center}
\vspace{-4.5cm}
\caption{The anisotropy parameter $a$ (see text) as a function of the 
pion-kinetic energy for the central event class. The circles represent the 
data and the squares indicate the result obtained by fitting the CBUU-results 
in the regime $0.0 \le cos \theta \le 0.9$ by $const(1 + a(E) cos^2\theta)$ for 
fixed pion kinetic energy.}
\label{fig_dataaofe}
\end{figure}
In fig. \ref{fig_dataaofe} we show $a(E)$ as a function of the pion-kinetic
energy. The result obained from the CBUU-model is depicted by the squares while  
the circles indicate the data. The overall functional form of the data is 
reproduced by the calculation except for the pronounced peak  
for pion kinetic energies of $0.25$ to $0.35$ GeV. The CBUU-model also 
shows the increase in anisotropy for energies up to $0.3$ GeV from zero to 
an $a(E)$ of $\approx 1.0$. Above these energies we observe a slight  decrease 
of the anisotropy coefficient $a(E)$. \\ 
In fig. \ref{fig_taps1} we finally compare our calculations to the more 
recent data on $\pi^0$-production in $Au + Au$ collisions at 1.0 AGeV
from the TAPS-collaboration. 
The solid histogram shows the result of our calculation, while the shaded 
areas indicate the statistical error of the calculation. The squares represent 
the data form \cite{vmetag94,vmetag96}. Here we find that our calculation 
is in good agreement with the experimental data, only slightly 
overestimating the spectrum in the region of $ p_T \approx 0.2$ GeV/c. In 
comparison to earlier calculations within the BUU-model of ref. \cite{mosel95} 
the low $p_T$-behaviour of the pion-spectrum has improved. In addition we are now able to reproduce the 
high $p_T$-data while the earlier calculation overestimated the high 
$p_T$-spectrum. The results obtained by Bass et al. within the IQMD-model \cite{bass2} 
are $\approx$ 20 \% higher than ours in the low $p_T$-region, while   
both calculations are well in line for the other $p_T$-regions. 
\begin{figure}[h]
\vspace{-1cm}
\begin{center}
\hspace{-18mm}{\epsfig{file=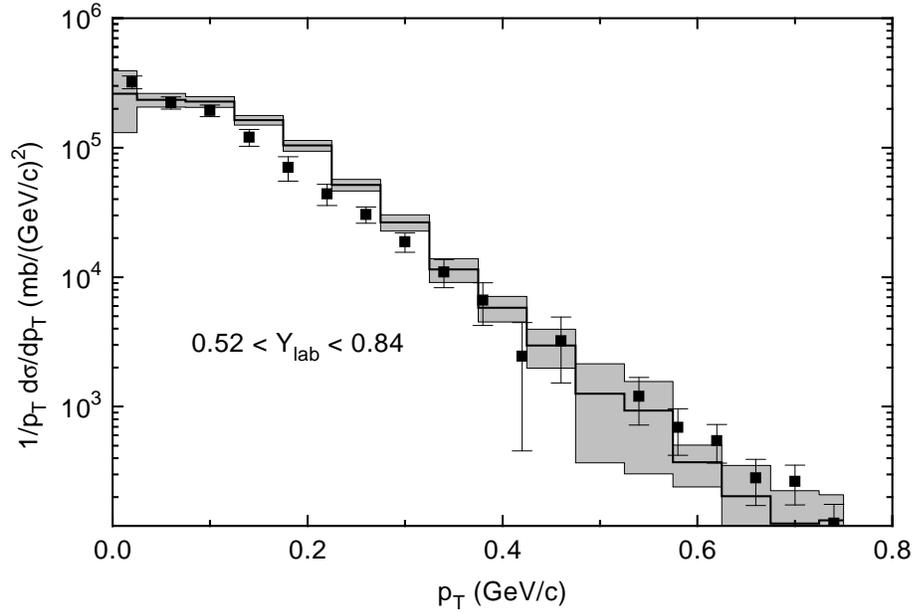,width=90mm}}
\vspace{-4.5cm}
\end{center}
\caption{Cross section for $\pi^0$-production in $Au + Au$ collisions  
at 1.0 GeV/A for $0.52 \le Y_{lab} \le 0.84$ as a function of the 
pion transverse momentum $p_T$. The solid histogram displays the result of 
the CBUU-calculation, while the shaded areas indicate the statistical 
error of the calculation. The data TAPS-data are represented  
by the squares. These data from ref. \protect\cite{vmetag94} have been renormalized by
a factor $0.6$ determined by a recent new analysis \protect\cite{vmetag96}.}
\label{fig_taps1}
\end{figure}

\end{section}
\begin{section}{Summary}
In this paper we have presented a new implementation of the
CBUU-transport-model for the description of relativistic heavy-ion collisions
up to energies of about $2$ GeV/A including baryonic resonances up to 
masses of $1.95$ $GeV/c^2$ as well as $\pi, \eta$ and $\rho$ mesons.
In addition to earlier implementations we also include 2$\pi$ production
channels via $NN \rightarrow \Delta \Delta \rightarrow N N \pi \pi$ and the
2$\pi$-decays of higher baryon resonances. The inclusive 1$\pi$ and 2$\pi$
cross sections from nucleon-nucleon collisions are found to be well reproduced
within our multi-resonance approach.

A detailed analysis of nucleus-nucleus collisions from 1 - 2 GeV/A shows that
the 2$\pi$ production channels increase the pion yield essentially in the
region of low transverse momenta $p_T$ while the one pion decays of the higher
baryon resonances are dominating the high $p_T$ or energetic part of the pion
spectrum. Thus the energetic part of the pion spectrum directly reflects
the abundancy of higher baryon resonances in the compressed stage of the
nucleus-nucleus collision.
  
In comparison to the 
$Ar + KCl$-data at $1.8$ AGeV for $\pi^-$-production
obtained at the BEVALAC we found an overall agreement of the spectral
$\pi^-$-distributions from the CBUU-model with the data except for the
low $p_T$-regime where the data are slightly underestimated. 
In comparing our calculations with the $\pi^0$-data of the
TAPS-collaboration for Au + Au at 1 GeV/A we find that the CBUU-model 
is in good agreement with the experimental spectrum, except for transverse momenta of 
$\approx 0.2$ GeV/c, where the data are overestimated. Furthermore, we have
shown that the experimentally observed angular anisotropies for pions 
are reproduced well and
can be understood in the framework of the multi-resonance picture. 
\end{section}

\end{document}